\documentclass[journal]{IEEEtran}

\usepackage{color}
\ifCLASSINFOpdf
  \usepackage[pdftex]{graphicx}
 \DeclareGraphicsExtensions{.pdf,.jpeg,.png}
\else
\fi
\usepackage[cmex10]{amsmath}
\usepackage{amsfonts}
\usepackage{cite}
\usepackage{url}

\newcommand{\JJ}{\jmath}

\begin{document}
\title{The Measurable Q Factor and Observable Energies of Radiating Structures}
\author{Miloslav~Capek,~\IEEEmembership{Student Member,~IEEE,}
        Lukas~Jelinek,
        Pavel~Hazdra,~\IEEEmembership{Member,~IEEE,}
        and~Jan~Eichler
\thanks{Manuscript received March 20, 2013; revised March 20, 2013.
This work was supported by the Grant Agency of the Czech Technical University in Prague, grant No. SGS12/142/OHK3/2T/13, and by the Grant Agency of the Czech Republic under project 13-09086S.}
\thanks{The authors are with the Department of Electromagnetic Field, Faculty of Electrical Engineering, Czech Technical University in Prague, Technicka 2, 16627, Prague, Czech Republic
(e-mail: miloslav.capek@fel.cvut.cz).}
}

%

\markboth{Journal of \LaTeX\ Class Files,~Vol.~6, No.~1, January~2007}%
{Capek \MakeLowercase{\textit{et al.}}: The Measurable Q Factor Based on Potentials}
%
\maketitle

\begin{abstract}
New expressions are derived to calculate the Q factor of a radiating device. The resulting relations link Q based on the frequency change of the input impedance at the input port ($Q_X$, $Q_Z$) with expressions based solely on the current distribution on an radiating device. The question of which energies of a radiating system are observable is reviewed, and then the proposed Q factor as defined in this paper is physical.
The derivation is based on potential theory rather than fields. This approach hence automatically eliminates all divergent integrals associated with electromagnetic energies in infinite space.
 
The new formulas allow us to study the radiation Q factor for antennas without feeding (through e.g. Characteristic Modes) as well as fed by an arbitrary number of ports.
The new technique can easily be implemented in any numerical software dealing with current densities. To present the merits of proposed technique, three canonical antennas are studied. Numerical examples show excellent agreement between the measurable $Q_Z$ derived from input impedance and the new expressions.

\end{abstract}


%
\IEEEpeerreviewmaketitle

\section{Introduction}
\label{Intro}
\IEEEPARstart{T}{he} radiation Q factor is recognized as one of the most significant parameters of the radiating system and its evaluation for antennas has long been discussed in the literature, see e.g. \cite{VolakisChenFujimoto_SmallAntennas_MiniatrurizTechniques} and references therein. The most recent approaches by Vandenbosch \cite{Vandenbosch_ReactiveEnergiesImpedanceAndQFactorOfRadiatingStructures} and Gustafsson \cite{Gustaffson_StoredElectromagneticEnergy_arXiv} use the actual distribution of the sources of radiation (currents) from which the electromagnetic energies and radiated power are evaluated. It has recently been shown \cite{GustafssonCismasuJonsson_PhysicalBoundsAndOptimalCurrentsOnAntennas_TAP} that Q as defined in \cite{Vandenbosch_ReactiveEnergiesImpedanceAndQFactorOfRadiatingStructures} (i.e. with the ``radiated energy" included) may deliver nonphysical negative values. Hence the question of which energies should be included as stored in the Q factor is still unsolved.

Rhodes \cite{Rhodes_ObservableStoredEnergiesOfElectromagneticSystems} poses exactly this question and develops formulas for the observable energies, i.e. energies which are measurable and thus physical. He defines the observable energy as that part of the total energy that has a measurable effect upon the input impedance and hence upon the frequency bandwidth. His results are interesting from the theoretical point of view, but since electric and magnetic fields in all the space are involved, they are not practical for numerical calculations.

It is known that the total energy of a radiating system in the frequency domain is infinite. This is true for the total energy evaluated from electromagnetic fields (which are stored in an infinite volume) \cite{Rhodes_ObservableStoredEnergiesOfElectromagneticSystems}, \cite{Rhodes_AReactanceTheorem}. Rhodes \cite{Rhodes_OnTheStoredEnergyOfPlanarApertures} showed that for observable energies the infinities in the integrals cancel in a special way, leaving a finite residue. Vandenbosch \cite{Vandenbosch_ReactiveEnergiesImpedanceAndQFactorOfRadiatingStructures} was able to analytically subtract the far-field energy from the total energy, isolating the residue and developing expressions for modified vacuum energies based on the currents at the radiating device, and he used them for evaluating $Q$.

This paper is inspired by \cite{Vandenbosch_ReactiveEnergiesImpedanceAndQFactorOfRadiatingStructures}, \cite{Gustaffson_StoredElectromagneticEnergy_arXiv}, \cite{Rhodes_ObservableStoredEnergiesOfElectromagneticSystems} and \cite{Rhodes_AReactanceTheorem}, but the line of reasoning is different. It is recognized here for the first time that the only useful and reasonable $Q$ factors of a radiator are the measurable ones, based on frequency changes of the input reactance $Q_X$, or more generally input impedance $Q_Z$, see \cite{YaghjianBest_ImpedanceBandwidthAndQOfAntennas}. The proposed development connects sources of radiation (surface currents flowing on an antenna) and the ``external world'', represented by the frequency behavior of the input impedance at the input port through the complex Poynting theorem. The necessary frequency derivatives on the source side are performed analytically at the level of electromagnetic potentials \cite{Jackson_ClassicalElectrodynamics}, which are advantageously utilized instead of field quantities \cite{Carpenter_ElectromagneticEnergyAndPowerInTermsOfChargesAndPotentialsInsteadOfFields}, \cite{HazdraCapekEichler_CommentsToGuy1}. Consequently, there are no infinite integrals present in the derivations. Similarly to previous works, we assume electric currents flowing in free space.

The main result is the expression for $Q_Z$ in terms of different electromagnetic quantities, linked to the current and charge on the antenna through three energy functionals arising from the complex Poynting theorem and its frequency differentation. In this way, a generalized impedance theorem for antennas is established, assuming not only frequency changes of Green's function, as in \cite{Vandenbosch_ReactiveEnergiesImpedanceAndQFactorOfRadiatingStructures} or \cite{Gustaffson_StoredElectromagneticEnergy_arXiv}, but also frequency changes of the current itself. As we show later in the paper, this gives a new additional term: the energy associated to reconfiguration of the current.

A huge advantage over $Q_Z$ as defined by Yaghjian and Best \cite{YaghjianBest_ImpedanceBandwidthAndQOfAntennas} is the possibility of using new expressions for modal currents (i.e. currents computed for a structure without any feeding, see also \cite{HarringtonMautz_TheoryOfCharacteristicModesForConductingBodies}, \cite{CapekHazdraEichler_AMethodForTheEvaluationOfRadiationQBasedOnModalApproach}). It is also possible to examine only a part of the structure of interest and to determine how much this part of the antenna affects the overall $Q$. In contrast with the quality factors derived in \cite{Vandenbosch_ReactiveEnergiesImpedanceAndQFactorOfRadiatingStructures} and \cite{Gustaffson_StoredElectromagneticEnergy_arXiv}, the $Q_Z$ proposed here is a measurable quantity and hence of interest for the design of arbitrary antennas with respect to their bandwidth.

The paper is organized as follows. In section \ref{theory}, the measurable $Q$ is derived in terms of the electric currents flowing on the antenna. Section \ref{discussion} discusses differences between previous attempts to calculate radiation $Q$ and the newly derived formulas. Section \ref{numStudies} presents numerical examples to verify the proposed theory on three representative antennas: a dipole, a loop and, a small double U-notched loop antenna. The consequences and applications are discussed, and selected results are compared with FEKO \cite{feko} and CST \cite{cst} software.

\section{Measurable Q-factor in terms of field sources}
\label{theory}
The purpose of the following derivations is to connect the measurable quality factor with the sources of the field. We will not a-priori rely on the classic expression $Q = \omega W/P$ as the defining relation with $W$ being the total reactive energy and $P$ the radiated power. Rather, we start with quality factor $Q_Z$, which originates from the behavior of the RLC circuit \cite{GustafssonNordebo_BandwidthQFactorAndResonanceModelsOfAntennas} and which has been shown to be useful also for estimating antenna performance regarding its impedance bandwidth \cite{YaghjianBest_ImpedanceBandwidthAndQOfAntennas}:
\begin{equation}
\label{EqX1}
Q_Z = \frac{\omega }{2 R_{\mathrm{in}}} \left|\frac{\partial Z_\mathrm{in}}{\partial \omega }\right|=\left| Q_R + \JJ Q_X \right|,
\end{equation}
where
\begin{equation}
\label{eqX2}
Q_R + \JJ Q_X = \frac{\omega }{2 R_{\mathrm{in}}} \frac{\partial \left( R_{\mathrm{in}} + \JJ X_{\mathrm{in}} \right)}{\partial \omega },
\end{equation}
$\jmath = \sqrt{-1}$ and $Z_{\mathrm{in}} = R_{\mathrm{in}} + \JJ X_{\mathrm{in}}$ is the input impedance of the antenna. In (\ref{eqX2}) and in the rest of the paper, time harmonic fields \cite{Harrington_TimeHarmonicElmagField} with angular frequency $\omega$ and the convention \mbox{${\boldsymbol{\mathcal{F}}}\left( t \right) = \sqrt{2} \Re \left\{ \mathbf{F} \left( \omega \right) {\mathrm{e}}^{\JJ \omega t} \right\}$} are assumed. 

In order to link (\ref{eqX2}) with the field sources and their energies, the power definition of the impedance is used:
\begin{equation}
\label{eqX3}
P_{\mathrm{in}} = \left( R_{\mathrm{in}} + \JJ X_{{\mathrm{in}}} \right) \left| I_0 \right|^2,
\end{equation}
where $P_{\mathrm{in}}$ is the complex power \cite{Jackson_ClassicalElectrodynamics}, and $I_0$ is the input current on the antenna's port. For the same situation, Poynting's theorem \cite{Jackson_ClassicalElectrodynamics} allows us to write
\begin{equation}
\label{eqX4}
P_{\mathrm{in}} =  - \int\limits_\Omega  \mathbf{E} \cdot \mathbf{J}^\ast \, \mathrm{d} \mathbf{r}  = \JJ \omega \int\limits_\Omega  \left( \mathbf{A} \cdot \mathbf{J}^* - \varphi \rho^* \right) \, \mathrm{d} \mathbf{r},
\end{equation}
where $\mathbf{E}$ is the electric field intensity, $\mathbf{J}$ is the current density, and $\rho$ is the charge density inside $\Omega$ region, respectively, $\mathbf{A}$ and $\varphi$ are the vector and scalar potential, respectively, and $\ast$ denotes complex conjugation. In the last step, the electromagnetic potentials \cite{Jackson_ClassicalElectrodynamics} have been used, see the appendix~\ref{appA}. Furthermore, using radiation integrals for $\mathbf{A} $, $\varphi $ in the Lorentz gauge and charge conservation, (\ref{eqX4}) can be rewritten as \cite{Jackson_ClassicalElectrodynamics}
\begin{equation}
\label{eqX5}
P_{\mathrm{in}} = \left( P_{\mathrm{m}} - P_{\mathrm{e}} \right) + \JJ \omega \left( W_{\mathrm{m}} - W_{\mathrm{e}} \right),
\end{equation}
where
\begin{subequations}
\begin{align}
\label{eqX6a}
W_{\mathrm{m}} - \JJ \frac{P_{\mathrm{m}}}{\omega} &= k^2 \mathcal{L} \left( \mathbf{J}, \mathbf{J} \right) , \\
\label{eqX6b}
W_{\mathrm{e}} - \JJ \frac{P_{\mathrm{e}}}{\omega} &= \mathcal{L} \left( \nabla \cdot \mathbf{J}, \nabla \cdot \mathbf{J} \right),
\end{align}
\end{subequations}
and where, in order to ease the notation, the following energy functional
\begin{equation}
\label{eqX7}
\mathcal{L} \left( \mathbf{U}, \mathbf{V} \right) = \frac{1}{4 \pi \epsilon \omega^2} \int\limits_{\Omega '} \int\limits_{\Omega} \mathbf{U} \left( \mathbf{r} \right) \cdot \mathbf{V}^\ast \left( \mathbf{r} ' \right) \frac{\mathrm{e}^{-\JJ k R}}{R} \, \mathrm{d} \mathbf{r} \, \mathrm{d} \mathbf{r} '
\end{equation}
has been defined. In (\ref{eqX7}), \mbox{$R = \left\| \mathbf{r} - \mathbf{r} ' \right\|$} is the Euclidean distance, \mbox{$k = \omega /{c_0}$} is the wavenumber and $c_0$ is the speed of light. The integration in principle runs over the entire space, but assuming sources of finite extent, the integrals are always finite. Within the chosen naming convention in (\ref{eqX5}), (\ref{eqX6a}), (\ref{eqX6b}), the quantity $W_{\mathrm{m}}$ is usually related to magnetic energy, while $W_{\mathrm{e}}$ is usually related to electric energy. This association is not unique, however, under the assumption of a linear, passive and lossless antenna, $P_{\mathrm{m}} - P_{\mathrm{e}}$ is the power radiated by the antenna and $ \omega \left( W_{\mathrm{m}} - W_{\mathrm{e}} \right)$ is the net reactive power. Assuming now input current $I_0 = 1\,\mathrm{A}$, substituting (\ref{eqX3}), (\ref{eqX5}) into (\ref{eqX2}), and comparing with (\ref{eqX6a}) and (\ref{eqX6b}), it is straightforward to arrive at\footnote{The upper index $(n)$ in the following expressions denotes the number of included $P$-terms and $W$-terms.}
\begin{equation}
\label{eqX8}
\begin{split}
& Q_R^{\left( 4 \right)} + \JJ Q_X^{\left( 4 \right)} = \frac{\omega }{2 \left(P_\mathrm{m}-P_\mathrm{e}\right)} \frac{\partial \Big(\left(P_{\mathrm{m}}-P_\mathrm{e}\right) + \JJ \omega \left(W_\mathrm{m}-W_\mathrm{e}\right)\Big)}{\partial \omega } \\
&= \frac{\left( P_{\mathrm{m}} + P_{\mathrm{e}} + P_{\mathrm{rad}} + P_{\omega}\right) + \JJ \omega \left( W_{\mathrm{m}} + W_{\mathrm{e}} + W_{\mathrm{rad}} + W_{\omega} \right)}{2 \left( P_{\mathrm{m}} - P_{\mathrm{e}} \right)},
\end{split}
\end{equation}
in which\footnote{Please note that the term denoted as $P_{\mathrm{rad}}$, is not radiated power.}
\begin{subequations}
\begin{align}
\label{eqX9a}
W_{\mathrm{rad}} - \JJ \frac{P_{\mathrm{rad}}}{\omega} &=  - \JJ k \left( k^2 \mathcal{L}_{\mathrm{rad}} \left( \mathbf{J}, \mathbf{J} \right) - \mathcal{L}_{\mathrm{rad}} \left( \nabla \cdot \mathbf{J}, \nabla \cdot \mathbf{J} \right) \right), \\
\label{eqX9b}
W_{\omega} - \jmath \frac{P_{\omega}}{\omega} &=  k^2 \mathcal{L}_{\omega} \left( \mathbf{J}, \mathbf{J} \right) - \mathcal{L}_{\omega} \left( \nabla \cdot \mathbf{J}, \nabla \cdot \mathbf{J} \right),
\end{align}
\end{subequations}
and where two more energy functionals are defined as
\begin{subequations}
\begin{align}
\label{eqX10a}
\mathcal{L}_{\mathrm{rad}} \left( \mathbf{U}, \mathbf{V} \right) &= \frac{1}{4 \pi \epsilon \omega^2} \int\limits_{\Omega '} \int\limits_{\Omega} \mathbf{U} \left( \mathbf{r} \right) \cdot \mathbf{V}^\ast \left( \mathbf{r} ' \right) \mathrm{e}^{-\JJ k R} \, \mathrm{d} \mathbf{r} \, \mathrm{d} \mathbf{r} ' , \\
\label{eqX10b}
\mathcal{L}_{\omega} \left( \mathbf{U}, \mathbf{V} \right) &= \frac{1}{4 \pi \epsilon \omega^2} \int\limits_{\Omega '} \int\limits_{\Omega} \omega \frac{\partial \mathbf{U} \left( \mathbf{r} \right) \cdot \mathbf{V}^\ast \left( \mathbf{r} ' \right)}{\partial\omega} \frac{\mathrm{e}^{-\JJ k R}}{R} \, \mathrm{d} \mathbf{r} \, \mathrm{d} \mathbf{r} ',
\end{align}
\end{subequations}
for details see the appendix~\ref{appB}. In (\ref{eqX8}), (\ref{eqX9a}) and (\ref{eqX9b}) the quantity $W_{\mathrm{rad}} - \JJ P_{\mathrm{rad}}/\omega$ can be attributed to the energy associated with radiation \cite{Vandenbosch_ReactiveEnergiesImpedanceAndQFactorOfRadiatingStructures}, \cite{Rhodes_ObservableStoredEnergiesOfElectromagneticSystems}, while the term $W_{\omega} - \JJ P_{\omega}/\omega$ should be interpreted as the energy needed for the current (charge) reconfiguration during a frequency change.

Neglecting the $W_{\omega}$ and $P_{\omega}$ terms in (\ref{eqX8}) results in
\begin{equation}
\label{eqX11}
Q_R^{\left( 3 \right)} + \JJ Q_X^{\left( 3 \right)} = \frac{\left( P_{\mathrm{m}} + P_{\mathrm{e}} + P_{\mathrm{rad}} \right) + \JJ \omega \left( W_{\mathrm{m}} + W_{\mathrm{e}} + W_{\mathrm{rad}} \right)}{2 \left( P_{\mathrm{m}} - P_{\mathrm{e}} \right)},
\end{equation}
which is just the radiation quality factor derived in \cite{Vandenbosch_ReactiveEnergiesImpedanceAndQFactorOfRadiatingStructures}, \cite{Gustaffson_StoredElectromagneticEnergy_arXiv}.

By omitting also the terms associated with radiation (which are usually small in comparison with the reactive power), one obtains
\begin{equation}
\label{eqX12}
Q_R^{\left( 2 \right)} + \JJ Q_X^{\left( 2 \right)} = \frac{\left( P_{\mathrm{m}} + P_{\mathrm{e}} \right) + \JJ \omega \left( W_{\mathrm{m}} + W_{\mathrm{e}} \right)}{2 \left( P_{\mathrm{m}} - P_{\mathrm{e}} \right)}.
\end{equation}
Note that $Q_X^{(2)}$ is the classical definition of the radiation quality factor \cite{CollinRotchild_EvaluationOfAntennaQ}.

\section{Discussion}
\label{discussion}
This section presents some important remarks:

\begin{itemize}
    \item The derivation of the general result (\ref{eqX8}) required only measurable quantities as radiated power $(P_\mathrm{m}-P_\mathrm{e})$ and net reactive power $\omega(W_\mathrm{m}-W_\mathrm{e})$. This is in contrast to the approach in \cite{Vandenbosch_ReactiveEnergiesImpedanceAndQFactorOfRadiatingStructures}, \cite{Gustaffson_StoredElectromagneticEnergy_arXiv}, which required the ambiguous separation of the net reactive power into an electric part and a magnetic part.
    \item The structure of the developed quality factor is compatible with the primary definition $Q = \omega W/P$, indirectly validating  (\ref{EqX1}) in \cite{YaghjianBest_ImpedanceBandwidthAndQOfAntennas}. It is now clearly seen how the change of input impedance is transferred into different forms of energy terms arising from various $\omega$ derivatives of (\ref{eqX3}).
    \item The $Q_Z^{(4)}$ expressed in (\ref{eqX8}) holds for any angular frequency $\omega$ and represents an untuned Q factor that has the strict physical meaning only in the self-resonances of the antenna $\Omega$. One can, however, compensate the nonzero reactive energy $W_\mathrm{m}-W_\mathrm{e}$ of the antena at each frequency by an additional energy $W_\mathrm{add}$ that is mostly concentrated in a adjacent tuning region\footnote{The same consideration as in \cite{YaghjianBest_ImpedanceBandwidthAndQOfAntennas}, Fig.~2 is made for the field concept.} $\Omega_\mathrm{add}$, containing currents $\mathbf{J}_\mathrm{add}$, so that the antenna system $\Omega_0 = \Omega \cup \Omega_\mathrm{add} $ is tuned to the resonance at $\omega_0$. Considering now that the tuning region is lossless and non-radiative, there is \mbox{$Q_R^{\left( 4 \right)} (\mathrm{tuned}) = Q_R^{\left( 4 \right)}$}. Furthermore, it is possible to calculate the $Q_X^{\left( 4 \right)} (\mathrm{tuned})$ according to (\ref{eqX8})
\begin{equation}
\label{eqDiss3}
Q_X^{\left( 4 \right)} (\mathrm{tuned}) = \frac{\omega_0 \left( W_{\mathrm{m}} + W_{\mathrm{e}} + W_{\mathrm{rad}} + W_\omega + W_{\mathrm{add}}\right)}{2 \left( P_{\mathrm{m}} - P_{\mathrm{e}} \right)},
\end{equation}
substituting $\mathbf{J}_0 = \mathbf{J} + \mathbf{J}_\mathrm{add}$ into the related functionals (\ref{eqX7}), (\ref{eqX10a}) and (\ref{eqX10b}). At this point it is important to note, that $W_\mathrm{add}$ is a function not only of $\mathbf{J}_\mathrm{add}$ but of $\mathbf{J}$ as well. This results from the fact that $W$-terms are bilinear forms and thus $W_\mathrm{add}$ consists of self-terms $\left( \mathbf{J}_\mathrm{add}, \mathbf{J}_\mathrm{add} \right)$ as well as cross-terms $\left( \mathbf{J}, \mathbf{J}_\mathrm{add} \right)$ and its permutation.
   \item In order to get an additional insight, imagine that the compensation is made by a serial lumped reactance $X_\mathrm{add} = - X_\mathrm{in}$. Using the circuit concept one obtains \mbox{$Q_X (\mathrm{tuned}) \propto \partial \left( X_\mathrm{add} + X_\mathrm{in} \right) / \partial \omega_0$}. On the contrary, the field concept leads to \mbox{$Q_X (\mathrm{tuned}) \propto \partial \, \omega_0 \left( W_\mathrm{m} (\mathbf{J}_0) - W_\mathrm{e} (\mathbf{J}_0) \right) / \partial\omega_0$}. Generaly, these two $Q_X (\mathrm{tuned})$ will differ, since the circuit concept neglects all the cross-terms mentioned in the previous point. In other words, $X_0 \neq X_\mathrm{in} + X_\mathrm{add}$, where $X_0$ is the measured input reactance of the antenna system (including the tuning region).
However, considering the compensation made by ideal lumped elements, the energy cross-terms will become negligible in comparison to the self-terms due to the field localization at the lumped circuits. In such cases we can aproximatelly write \mbox{$W_\mathrm{add} \approx L_\mathrm{add} \Leftrightarrow W_\mathrm{m} < W_\mathrm{e}$} in the case of added serial inductor, and \mbox{$W_\mathrm{add} \approx 1/ \omega_0^2 C_\mathrm{add} \Leftrightarrow W_\mathrm{m} > W_\mathrm{e}$} in the case of added serial capacitor.
    \item On the basis of the previous discussion, the classical concept of tuned Q can be adopted into the proposed definition. Considering that the energies $W_{\mathrm{m}}$ and $W_{\mathrm{e}}$ are positively semi-definite, we obtain from (\ref{eqDiss3})
\begin{equation}
\label{eqDiss4}
Q_X^{\left( 4 \right)} (\mathrm{tuned}) = \frac{2 \omega_0 \max\left\{ W_{\mathrm{m}}, W_{\mathrm{e}}\right\} + \omega_0 \left( W_{\mathrm{rad}} + W_\omega \right)}{2 \left( P_{\mathrm{m}} - P_{\mathrm{e}} \right)},
\end{equation}
or neglecting the $W_\mathrm{rad}$ and $W_\omega$
\begin{equation}
\label{eqDiss5}
Q_X^{\left( 2 \right)} (\mathrm{tuned}) = \frac{\omega_0 \max\left\{ W_{\mathrm{m}}, W_{\mathrm{e}}\right\}}{\left( P_{\mathrm{m}} - P_{\mathrm{e}} \right)},
\end{equation}
which is the classical definition of tuned Q, \cite{CollinRotchild_EvaluationOfAntennaQ}, \cite{Vandenbosch_ReactiveEnergiesImpedanceAndQFactorOfRadiatingStructures}.
    \item The only difference between (\ref{eqX8}) and the quality factor derived in \cite{Vandenbosch_ReactiveEnergiesImpedanceAndQFactorOfRadiatingStructures}, \cite{Gustaffson_StoredElectromagneticEnergy_arXiv} is the presence of $W_{\omega}$, $P_{\omega}$ terms. However, they are not in practice observable in $Q_Z$ for the cases in this paper, as will be shown in the next section. Furthermore, the terms $W_{\omega}$, $P_{\omega}$, $W_{\mathrm{rad}}$, $P_{\mathrm{rad}}$ cannot be strictly separated from each other, as their (internal) energy exchange cannot be detected at the port.
    \item Only $W_{\omega}$, $P_{\omega}$ terms require current normalization (i.e. specification of the input current $I_0$). Dropping them (which fortunately has a very small effect on the measurable $Q_Z$ factor) thus allows the calculation of the $Q_Z$ factor of the arbitrary current distribution (for example modal currents) without referring to a particular feeding network.
    \item By analogy with \cite{Vandenbosch_ReactiveEnergiesImpedanceAndQFactorOfRadiatingStructures} and \cite{CapekHazdraEichler_AMethodForTheEvaluationOfRadiationQBasedOnModalApproach}, the derived expressions for the quality factor are easy to implement in any method of moment (MoM) \cite{Harrington_FieldComputationByMoM} code as a post processing routine. The only complication is the existence of $\cos (k R) / R$ terms in the energy functionals (\ref{eqX7}), (\ref{eqX10a}), (\ref{eqX10b}). These singularities are however removable and integrable analytically \cite{ArcioniBressanPerregrini_OnTheEvaluationOfTheDoubleSurfaceIntegralsArisingInTheApplicationOfETC}, \cite{CapekHazdraEichler_AMethodForTheEvaluationOfRadiationQBasedOnModalApproach}.
\end{itemize}

\section{Numerical results}
\label{numStudies}
In this section we will show numerical results for three canonical antennas, discuss the most important features of the Q factor defined by (\ref{eqX8}), and compare it with other available definitions. To this point, the expressions given in Sec.~\ref{theory} were implemented in our in-house MoM solver \cite{CapekHamouzHazdraEichler_ImplementationOfTCMinMatlab} based on RWG basis functions \cite{RaoWiltonGlisson_ElectromagneticScatteringBySurfacesOfArbitraryShape} in Matlab \cite{matlab}. Thanks to implementation on a GPU card \cite{CapekHazdraEichlerHamouzMazanek_AccelerationTechniquesInMatlab}, all the calculations are extremely fast (about 0.01\,s for one frequency sample and 200 RWG functions). Note that in order to keep the discussion as general as possible, a dimensionless quantity $ka$ is used instead of frequency, with $a$ being the smallest radius of a sphere circumscribing all the sources.

\subsection{A thin-strip dipole}
\label{Example1}
\begin{figure}[!t]
\centering
\includegraphics[width=9.2cm]{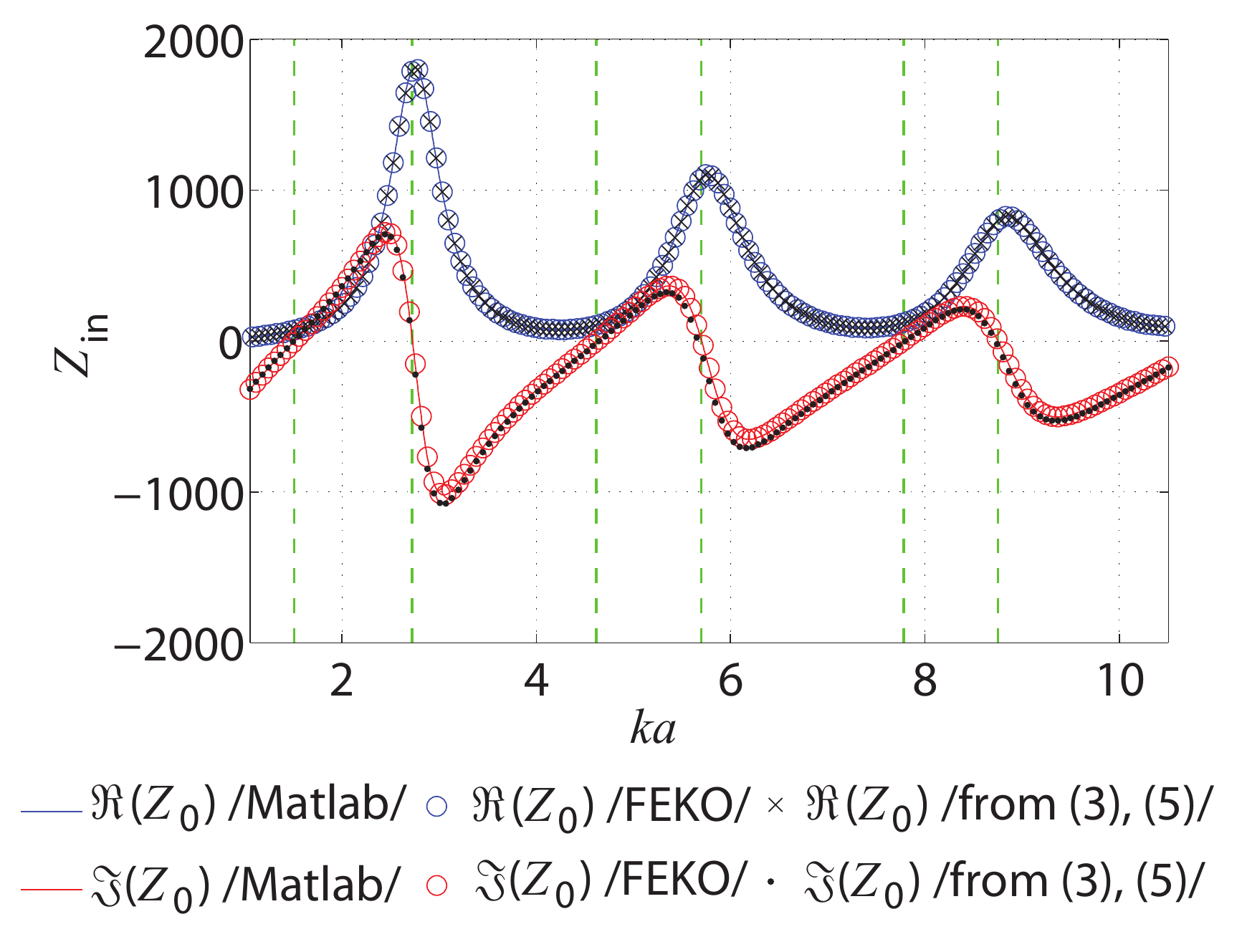}
\caption{Real and imaginary parts of the input impedance of a thin-strip dipole simulated in Matlab RWG-MoM and in FEKO software. The ``Matlab" and ``FEKO" parts are directly calculated as a ratio of voltage and current at the feeding port, while the part denoted as ``from (3) and (5)" comes from the direct integration of current distribution on the antenna. The green dashed lines mark resonances and antiresonances.}
\label{fig_dipole_fig1}
\end{figure}

The first example deals with a dipole radiating in free space. The dipole is made of an infinitesimally thin perfectly conducting strip with length $2L$ and width $w = 2L/250$. The dipole is discretized into 201 triangles and is fed by a delta gap \cite{Balanis_AdvancedTheory} in its center (the voltage corresponds to the input current $I_0 = 1\,\mathrm{A}$). The real and imaginary parts of the input impedance are shown in Fig.~\ref{fig_dipole_fig1}. For comparison, the dipole was also simulated in FEKO software. Note, that almost exact correspondence in Fig.~\ref{fig_dipole_fig1} validates the correct implementation of the MoM and the integration routines, and furthermore, it demonstrates the equality between (\ref{eqX3}) and (\ref{eqX5}). Good correspondence between the results justifies the use of our Matlab RWG-MoM code in the rest of the paper.

\begin{figure}[!t]
\centering
\includegraphics[width=9.2cm]{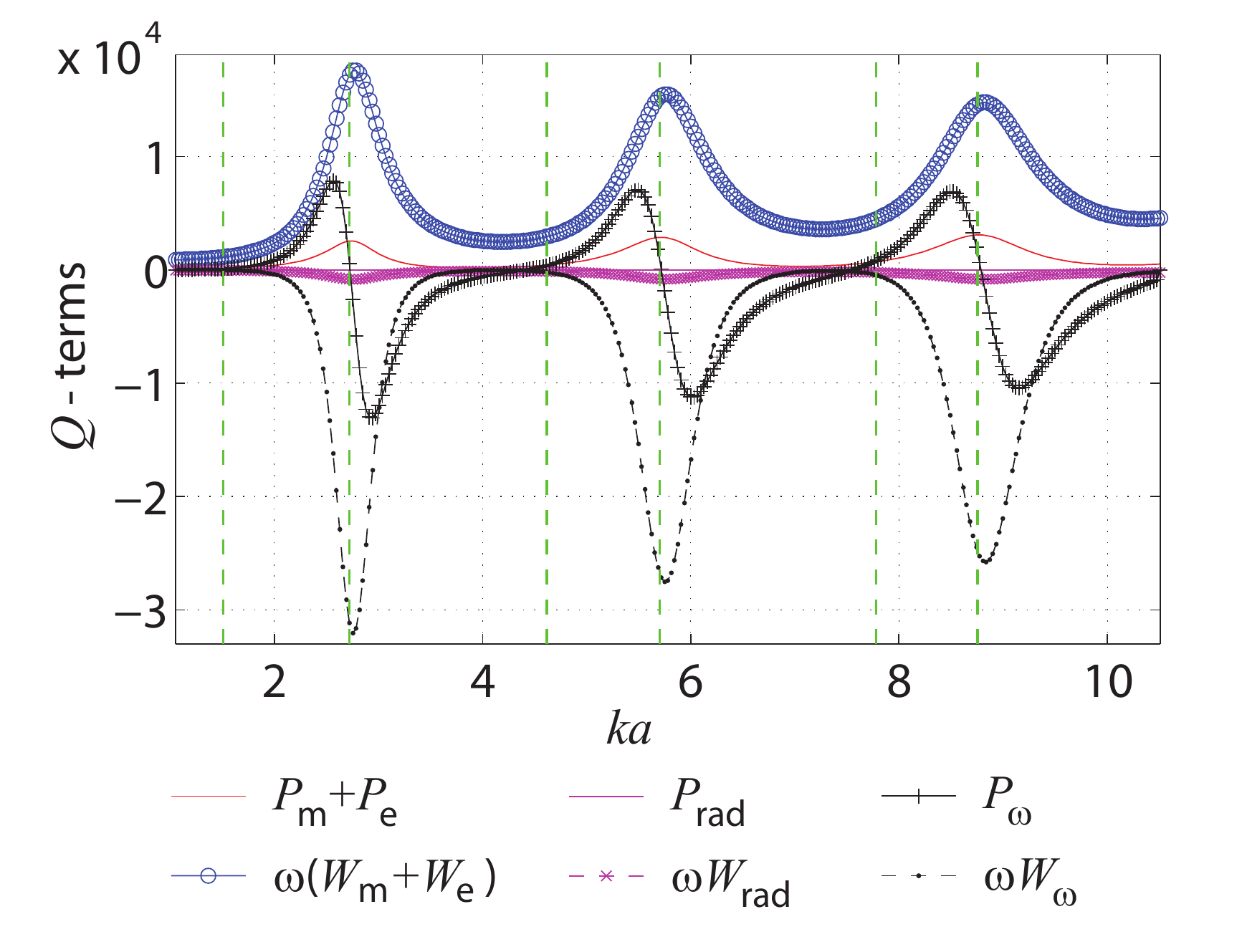}
\caption{The frequency dependence of terms composing the nominator of (\ref{eqX8}) for the thin-strip dipole of Fig.~\ref{fig_dipole_fig1}.}
\label{fig_dipole_fig5}
\end{figure}

We now turn to a brief discussion of the terms composing the nominator of (\ref{eqX8}). The most relevant terms are depicted in Fig.~\ref{fig_dipole_fig5}.
The first observation is that $P_{\mathrm{rad}}$ can be safely neglected. Its small value is caused by almost exact cancellation of the real parts of otherwise important terms $k^2 \mathcal{L}_{\mathrm{rad}} \left( \mathbf{J}, \mathbf{J} \right)$, $\mathcal{L}_{\mathrm{rad}} \left( \nabla \cdot \mathbf{J} , \nabla \cdot \mathbf{J} \right)$, see (\ref{eqX9a}). The same is approximately valid also for the $W_{\mathrm{rad}}$ term, though there the cancellation is not as perfect. There should thus be no important difference between the quality factor defined by (\ref{eqX11}) and by (\ref{eqX12}). The second observation relates to $W_{\omega} - \JJ P_{\omega}/\omega$. The absolute value of this quantity evidently reaches its maximum at antiresonances and its minimum in the vicinity of resonances. This is coherent with the interpretation as reconfiguration energy, mentioned in Sec.~\ref{theory}: stable eigenmodes exist in the vicinity of resonances, while the change from one eigenmode to another peaks at antiresonances.

\begin{figure}[!t]
\centering
\includegraphics[width=9.2cm]{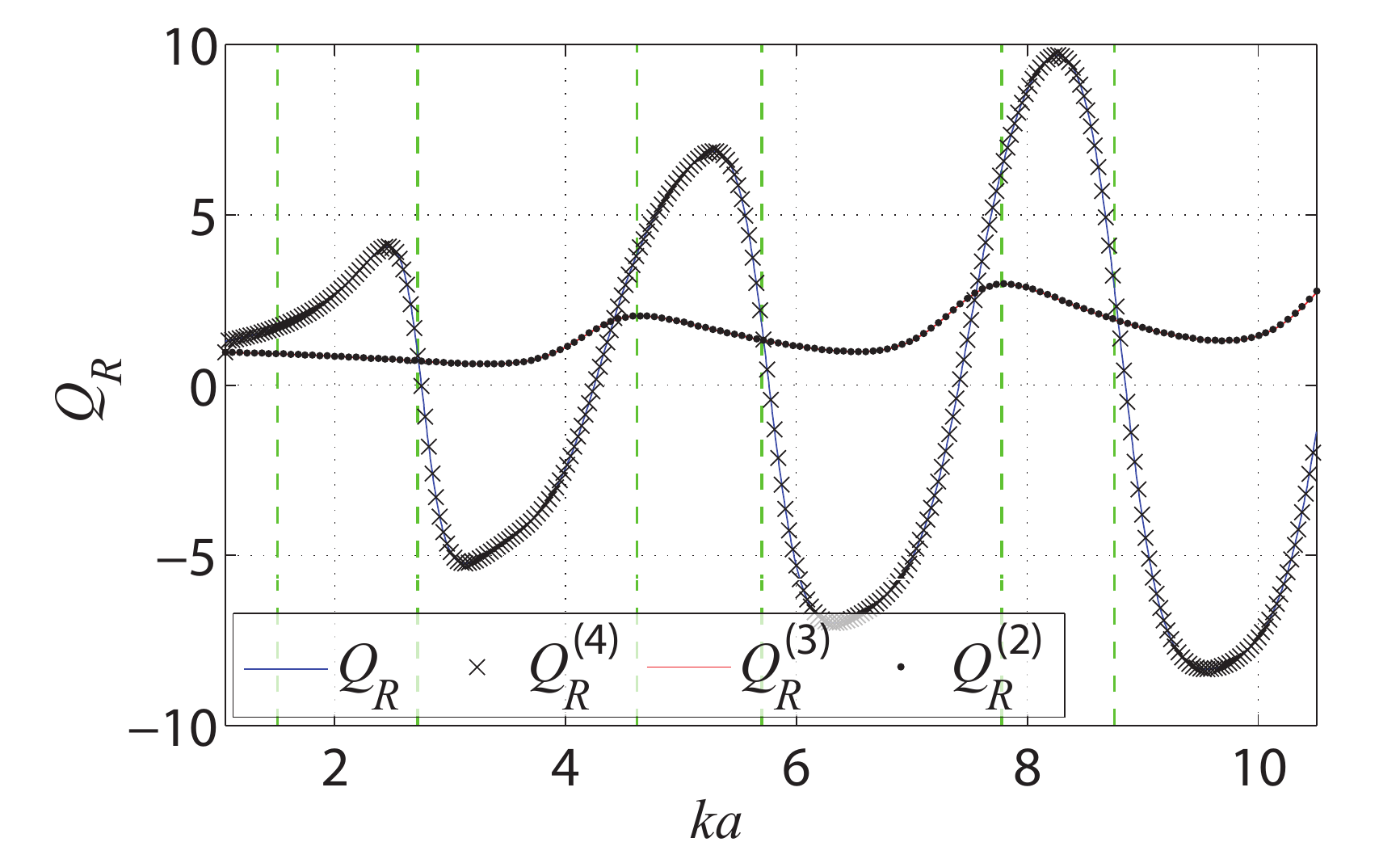}
\caption{Comparison of the radiation $Q_R$ factors of the thin-strip dipole of Fig.~\ref{fig_dipole_fig1}. The green dashed lines mark resonances and antiresonances, see Fig.~\ref{fig_dipole_fig1}.}
\label{fig_dipole_fig2}
\end{figure}

\begin{figure}[!t]
\centering
\includegraphics[width=9.2cm]{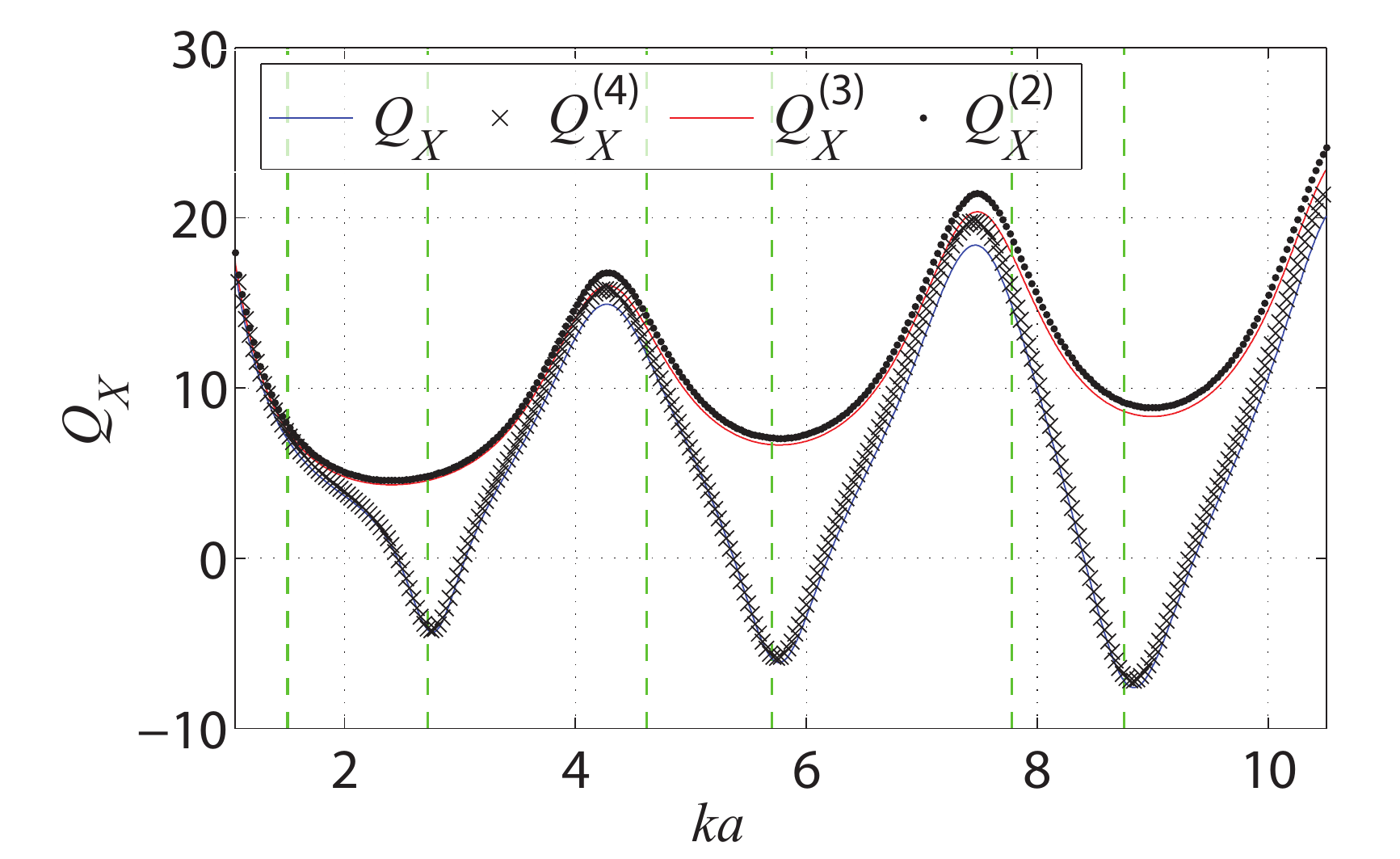}
\caption{Comparison of the radiation $Q_X$ factors of the thin-strip dipole of Fig.~\ref{fig_dipole_fig1}. The green dashed lines mark resonances and antiresonances, see Fig.~\ref{fig_dipole_fig1}.}
\label{fig_dipole_fig3}
\end{figure}

\begin{figure}[!t]
\centering
\includegraphics[width=9.2cm]{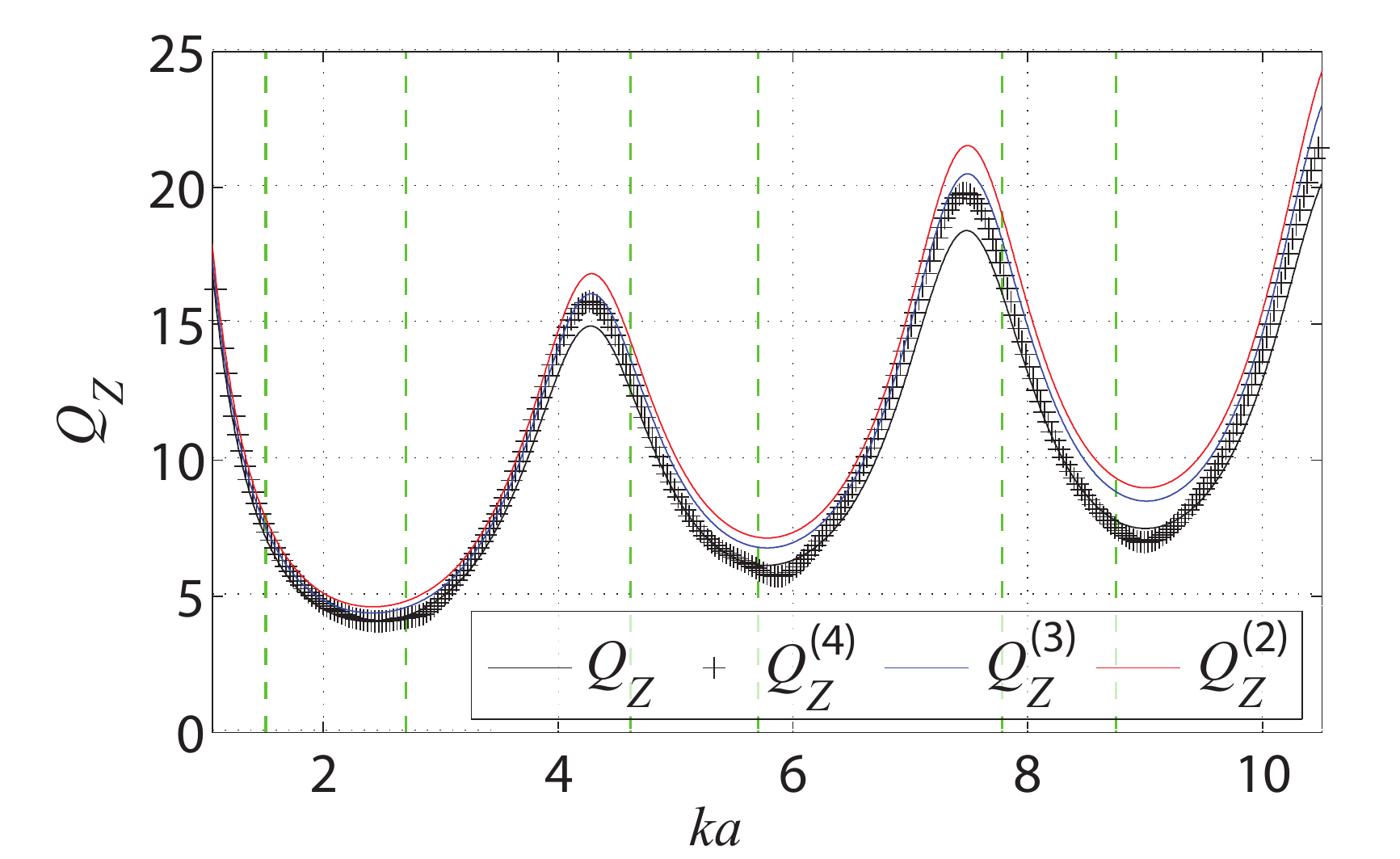}
\caption{Comparison of the radiation $Q_Z$ factors of the thin-strip dipole of Fig.~\ref{fig_dipole_fig1}. The green dashed lines mark resonances and antiresonances, see Fig.~\ref{fig_dipole_fig1}.}
\label{fig_dipole_fig4}
\end{figure}

In order to check the discussion above, the radiation quality factors given by (\ref{eqX2}), (\ref{eqX8}), (\ref{eqX11}) and (\ref{eqX12}) are depicted in Fig.~\ref{fig_dipole_fig2},~\ref{fig_dipole_fig3},~\ref{fig_dipole_fig4}. A central difference has been used to calculate (\ref{eqX2}). Note that the correspondence between (\ref{eqX2}) and (\ref{eqX8}) verifies the numerical implementation, since the expressions are analytically equal. As expected, the quality factors given by (\ref{eqX11}) and (\ref{eqX12}) are mostly alike at all frequencies. By contrast, the biggest difference between the $Q^{(4)}_R$, $Q^{(4)}_X$ factors given by (\ref{eqX8}) and $Q^{(3)}_R$, $Q^{(3)}_X$ and $Q^{(2)}_R$, $Q^{(2)}_X$ appears at antiresonances, which is due to the presence of $W_{\omega}$, $P_{\omega}$ terms. On the other hand, in the case of resonances the reconfiguration energy is small and all the depicted quality factors are very similar. Observing the differences in the $Q_R$, $Q_X$ factors, it is however quite remarkable that in the case of $Q_Z$ the $W_{\omega}$, $P_{\omega}$ terms play almost no role.

\subsection{Modal solution of a loop, and an analogy with two dipoles}
\label{Example2}

The second example reveals other benefits of the new technique: the utilization of modal methods. The former $Q_Z$ definition cannot be used in these cases. From the previous example we know that the $W_{\omega}$ term is important for calculating $Q_R$ and $Q_X$ but it can be omitted in calculating $Q_Z$. Thus, current normalization is not necessary, and no port needs to be specified. 

In this example, two basic radiators, a loop and two closely spaced semicircular dipoles that occupy the same volume as the loop does, were decomposed to the characteristic modes, \cite{HarringtonMautz_TheoryOfCharacteristicModesForConductingBodies}, \cite{CapekHazdraEichler_AMethodForTheEvaluationOfRadiationQBasedOnModalApproach}. The radius of the loop is $R$, and the length of the dipoles is $\pi R$. An infinitesimally thin perfectly conducting strip of width $R/12$ is considered both for the loop and the dipoles. The dipoles are separated by a gap of width $R/16$.

\begin{figure}[!t]
\centering
\includegraphics[width=9.2cm]{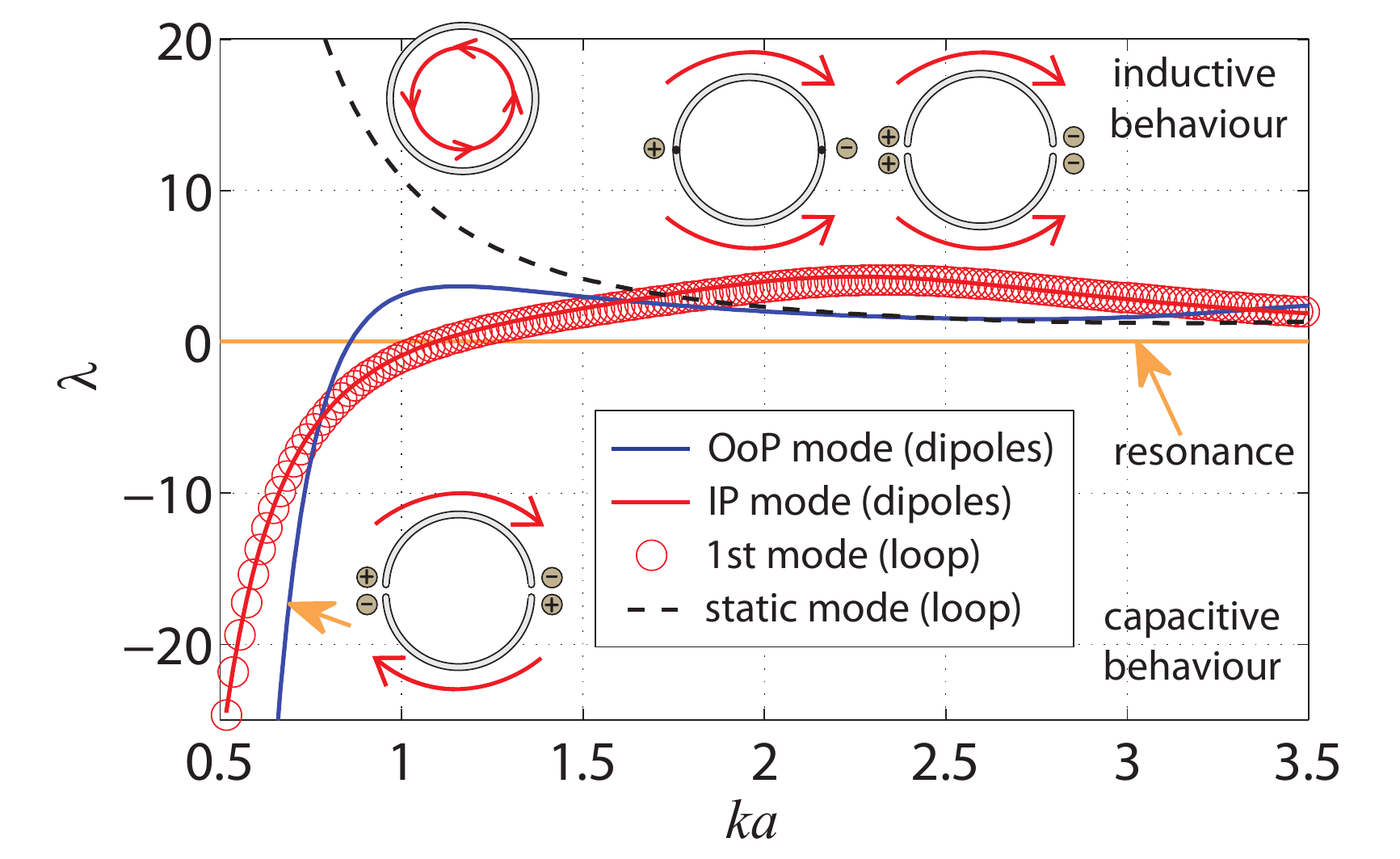}
\caption{The eigennumbers of two dipoles and the loop, IP stands for in-phase mode, OoP stands for out-of-phase mode.}
\label{fig_dipoles_fig7}
\end{figure}

The two dipole scenario consists of two possible modes around the first resonant frequency: the in-phase mode (IP) and the out-of-phase mode (OoP), see Fig.~\ref{fig_dipoles_fig7}. The solution of the loop at the same frequency consists of the static (inductive) mode and the first mode, as depicted in Fig.~\ref{fig_dipoles_fig7}. The eigennumbers $\lambda$ determine the modal behaviour, mode is capacitive for $\lambda <0$, inductive for $\lambda > 0$, and is in resonance for $\lambda = 0$. The eigennumbers for the loop and the dipoles are shown in Fig.~\ref{fig_dipoles_fig7}.
The $Q_Z$ factors defined by (\ref{eqX8}) were calculated from the modal currents. Thanks to the freedom in the current definition, we also calculated the case of OoP dipoles with the charge completely eliminated (setting $\nabla\cdot\mathbf{J} \equiv 0$).
\begin{figure}[!t]
\centering
\includegraphics[width=9.2cm]{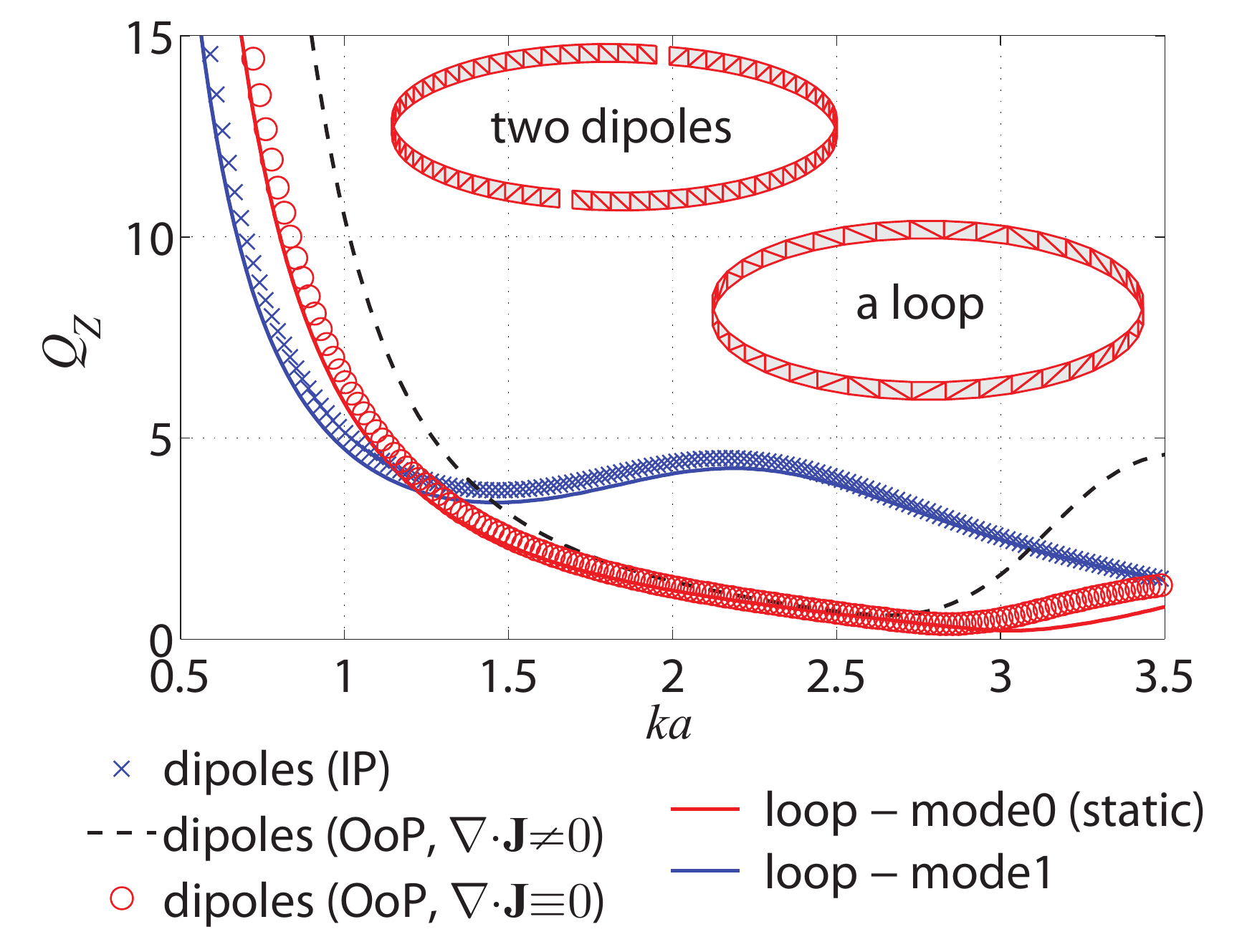}
\caption{Equivalence of two topologically different structures -- the loop and two closely spaced circular dipoles, IP stands for the in-phase mode, OoP stands for the out-of-phase mode.}
\label{fig_dipoles_fig8}
\end{figure}
As depicted in Fig.~\ref{fig_dipoles_fig8}, the $Q_Z$ of the static mode of the loop looks like the $Q_Z$ of the out-of-phase mode of semicircular dipoles with all charge terms eliminated, see the red line and the red circular marks in Fig.~\ref{fig_dipoles_fig8}. Similarly, the first mode of the loop has the same $Q_Z$ factor as the in-phase mode of the dipoles (no modification is needed in this case because the charge distribution is the same for both cases). Note that the static mode is always excited (irrespective of feed position). Thus, the static mode increases the total $Q$ at all frequencies \cite{CapekHazdraEichler_AMethodForTheEvaluationOfRadiationQBasedOnModalApproach}.

\subsection{Small U-notched loop antenna}
\label{Example3}

\begin{figure}[!t]
\centering
\includegraphics[width=9.2cm]{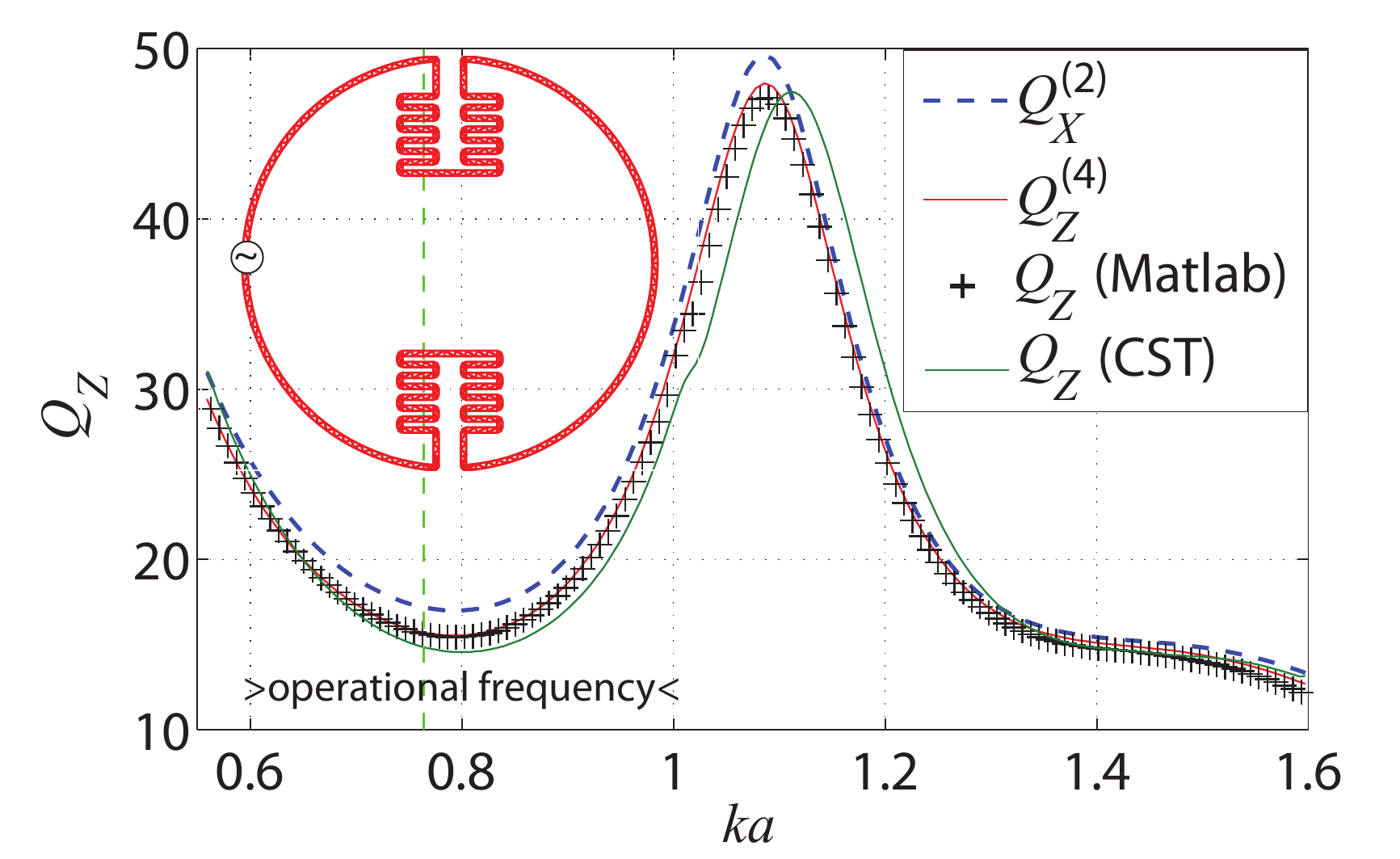}
\caption{Comparison of the $Q_Z$ factors from Matlab RWG-MoM (both definition (\ref{EqX1}) and definition (\ref{eqX8}) definitions are shown) and CST-MWS (definition (\ref{EqX1})). The antenna operates around the vertical dashed green line.}
\label{fig_dipole_fig11}
\end{figure}

The electrically small U-notched loop antenna was designed in CST-MWS \cite{cst}. The radius of the antenna is $R$, the width of the infinitesimally thin strip is $R/36$, and PEC is considered, see Fig.~\ref{fig_dipole_fig11}. To make the antenna electrically smaller, the parts with negligible current density are meandered. The same structure was simulated in Matlab RWG-MoM and decomposed into characteristic modes.


We can estimate the overall $Q_Z$ factor of the fabricated antenna approximately from the $-3\,\mathrm{dB}$ fractional bandwidth ($\mathrm{FBW}_{-3\, \mathrm{dB}}$) as \cite{YaghjianBest_ImpedanceBandwidthAndQOfAntennas}
\begin{equation}
\label{Derivation_Ex3_Eq3}
Q_Z \approx \frac{\sqrt{2}}{2} \frac{1}{\mathrm{FBW}_\mathrm{-3\,dB}}.
\end{equation}
Relation (\ref{Derivation_Ex3_Eq3}) holds for $Q \gg 1$. In the case of the manufactured U-notched antenna,
the Q factor (\ref{Derivation_Ex3_Eq3}) is equal to $14.8$. For the same procedure in CST-MWS we obtained $Q = 15.5$. From differentiation of $Z_\mathrm{in}$ in CST-MWS we obtained $Q_Z = 14.8$, and from integrating the current distribution in Matlab we obtained $Q_Z^{(4)} = 15.6$, see Fig.~\ref{fig_dipole_fig11}. The moderate difference between CST and Matlab in Fig.~\ref{fig_dipole_fig11} can be attributed to numerical issues.

\begin{figure}[!t]
\centering
\includegraphics[width=9.2cm]{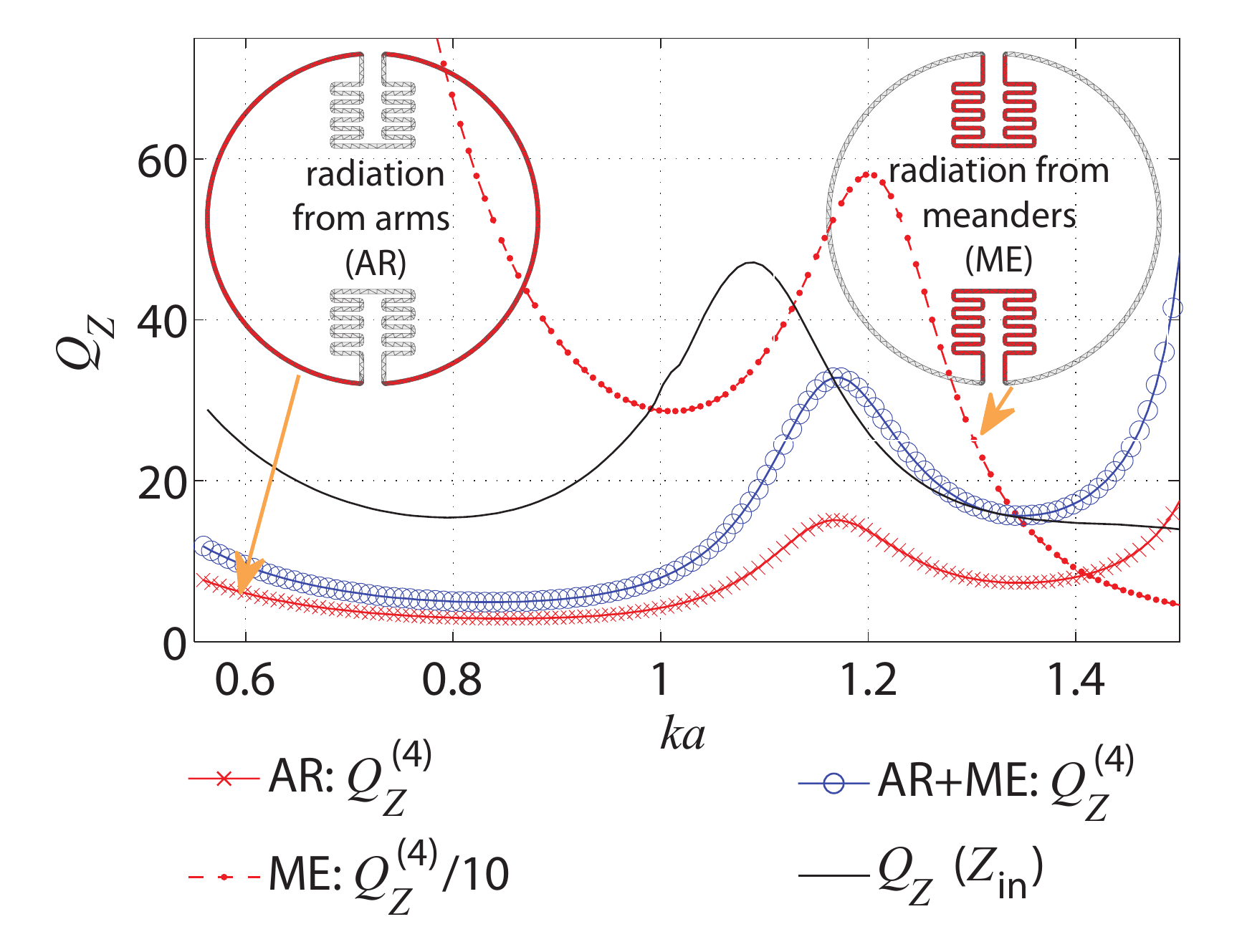}
\caption{A study of the particular $Q$ that corresponds to the selected radiating parts of the U-notched loop antenna. AR stands for radiating from the arms, ME stands for radiating from the meanders. The overall $Q_Z$ (solid black line) is added as a reference.}
\label{fig_dipole_fig12}
\end{figure}

The last discussed feature of the proposed definition is calculating the radiation $Q_Z$ factor for a selected part of a radiating device only. While the arms of the antenna radiate well, the meanders accumulate a great deal of net reactive power because of the out-of-phase currents. Thus, we try to calculate the $Q_Z$ of these two parts separately. To do so, the total current distribution $\mathbf{J}$ is separated as
\begin{subequations}
\begin{align}
\label{eq3_1A}
\mathbf{J}_1 (\mathbf{r}) = \mathbf{J} (\mathbf{r}) \delta_1 (\mathbf{r}),  \\
\label{eq3_1B}
\mathbf{J}_2 (\mathbf{r}) = \mathbf{J} (\mathbf{r}) \delta_2 (\mathbf{r}),
\end{align}
\end{subequations}
where $\delta_1 (\mathbf{r}) = 1$ for all $\mathbf{r}$ where $|\mathbf{r}| > 35R/36$ and $\delta_1 (\mathbf{r}) = 0$ otherwise, $\delta_2= 1-\delta_1$. The results are depicted in Fig.~\ref{fig_dipole_fig12}. While the particular $Q$ that corresponds to the arms of the antenna is very low, the Q factor corresponding to the meanders is extremely high (note that the corresponding values of $Q_Z$ are divided by 10 in Fig.~\ref{fig_dipole_fig12}).

If we sum up the energetic terms corresponding to the separated arms ($\mathbf{J}_1$) and meanders ($\mathbf{J}_2$) and calculate $Q_Z$, we do not get the overall $Q_Z$ of whole structure ($\mathbf{J}$), Fig.~\ref{fig_dipole_fig12}. This was expected, and the reason lies in the fact that the operators (\ref{eqX7}), (\ref{eqX10a}) and (\ref{eqX10b}) are not linear for $\mathbf{J} = \mathbf{J}_1 + \mathbf{J}_2$ and thus all possible interactions of the separated parts are omitted. However, these interactions can be calculated by substituting $\mathcal{L} \left( \mathbf{J}_1, \mathbf{J}_2 \right)$, $\mathcal{L} \left( \nabla\cdot\mathbf{J}_1, \nabla\cdot\mathbf{J}_2 \right)$ and similarly for (\ref{eqX10a}), (\ref{eqX10b}) into the $Q_Z$ calculation.

\section{Conclusion}
\label{concl}
A new formulation of the radiation Q factor is derived in terms of field sources instead of fields. The utilization of the complex Poynting theorem and potential theory had two main effects: a) interpretation and justification of the questionable concept of separating electric and magnetic energy in non-stationary fields is not required, b) integrations over the entire space are not present.

It is well known as the reactance theorem that the total energy of a passive electromagnetic system is proportional to the change of input impedance with frequency. The same derivation is analytically performed on the source side of the complex Poynting theorem, resulting in energetic expressions of a different nature, compactly expressed as functionals of the current. They form the observable quantities which can be measured through the input impedance -- and this is the only concept that can be physically tested and thus is of practical interest. Moreover it is shown that the formulas are also valid for modal currents, where no feeding is present.
 
A novel energy term, related to the current reshaping, is shown to be the cause of negative values of measured $Q_X$ in the antiresonances. Interestingly, this reconfiguration energy is almost not transferred into $Q_Z$.

The examples, presented here have verified the new expressions and have illustrated some benefits of the method. The proposed concept is easy to implement and offers new challenges in small antenna and MIMO antenna design, especially in conjunction with modal decomposition and optimization.
\appendices
\section{The Complex Power in Terms of Potentials}
\label{appA}
The purpose of this appendix is to derive the relation 
\begin{equation}
\label{EqA1}
- \int\limits_\Omega  {{\mathbf{E}} \cdot {{\bf{J}}^*}} \, {\mathrm{d}}{\mathbf{r}} = \jmath \omega \int\limits_\Omega  {\left( {{\mathbf{A}} \cdot {{\mathbf{J}}^*} - \varphi {\rho ^*}} \right)} \, {\mathrm{d}}{\mathbf{r}}
\end{equation}
which has been used in (\ref{eqX4}). The equality (\ref{EqA1}) is most easily derived by direct substitution of the defining relation of electromagnetic potentials \cite{Jackson_ClassicalElectrodynamics}
\begin{equation}
\label{EqA2}
{\mathbf{E}} =  - \nabla \varphi  - \jmath \omega {\mathbf{A}}
\end{equation}
into the LHS of (\ref{EqA1}). This leads to
\begin{equation}
\label{EqA3}
- \int\limits_\Omega  {{\mathbf{E}} \cdot {{\mathbf{J}}^\ast}} \, {\mathrm{d}}{\mathbf{r}} = \jmath \omega \int\limits_\Omega  {{\mathbf{A}} \cdot {{\mathbf{J}}^\ast}} \, {\mathrm{d}}{\mathbf{r}} + \int\limits_\Omega  {\nabla \varphi  \cdot {{\mathbf{J}}^\ast}} \, {\mathrm{d}}{\mathbf{r}}.
\end{equation}
The relation (\ref{EqA3}) can be further rewritten with the use of vector identity
\begin{equation}
\label{EqA4}
\nabla \varphi  \cdot {{\mathbf{J}}^\ast} = \nabla  \cdot \left( {{{\mathbf{J}}^\ast}\varphi } \right) - \varphi {\nabla  \cdot {{\mathbf{J}}^\ast}},
\end{equation}
continuity equation
\begin{equation}
\label{EqA5}
\nabla  \cdot {\bf{J}} =  - \jmath \omega \rho,
\end{equation}
and Gauss theorem into
\begin{equation}
\label{EqA6}
- \int\limits_\Omega  {{\mathbf{E}} \cdot {{\mathbf{J}}^\ast}} {\mathrm{d}}{\mathbf{r}} = \jmath \omega \int\limits_\Omega {\left( {{\mathbf{A}} \cdot {{\mathbf{J}}^\ast} - \varphi {\rho^\ast}} \right)} {\mathrm{d}}{\mathbf{r}} + \int\limits_{\partial \Omega } \varphi  {{\mathbf{J}}^\ast} \cdot {\mathrm{d}}{\mathbf{S}}.
\end{equation}
Now, using the fact that the current component normal to the surface $\partial\Omega$ is zero, the last term in (\ref{EqA6}) identically vanishes and we arrive at (\ref{EqA1}).

\section{Derivation of Relation (\ref{eqX8})}
\label{appB}

The first step of the derivation is the use of the radiation integrals for vector and scalar potentials in homogenous, isotropic and open region \cite{Jackson_ClassicalElectrodynamics}
\begin{equation}
\label{app_a1}
\mathbf{A} \left( \mathbf{r} \right) = \frac{\mu}{4 \pi} \int\limits_{\Omega '} \mathbf{J} \left( \mathbf{r} ' \right) \frac{\mathrm{e}^{-\JJ k R}}{R} \, \mathrm{d} \mathbf{r} '
\end{equation}
and
\begin{equation}
\label{app_a2}
\varphi \left( \mathbf{r} \right) = \frac{1}{4 \pi \epsilon} \int\limits_{\Omega '} \rho \left( \mathbf{r} ' \right) \frac{\mathrm{e}^{-\JJ k R}}{R} \, \mathrm{d} \mathbf{r} ',
\end{equation}
together with (\ref{EqA5}) and (\ref{eqX7}) to obtain
\begin{equation}
\label{app_b1}
\int\limits_\Omega  \left( \mathbf{A} \cdot \mathbf{J}^\ast - \varphi \rho^\ast \right) \, \mathrm{d} \mathbf{r} = k^2 \mathcal{L} \left( \mathbf{J},\mathbf{J} \right) - \mathcal{L} \left( \nabla \cdot \mathbf{J}, \nabla \cdot \bf{J} \right).
\end{equation}
The next step consists of substituing (\ref{eqX4}) into (\ref{eqX3}) and afterwards to (\ref{eqX2}) and evaluating various $\omega$ derivatives. In particular, there is
\begin{equation}
\label{app_b2}
\begin{split}
\displaystyle\frac{\partial {k^2} \mathcal{L} \left( \mathbf{J},\mathbf{J} \right)}{\partial \omega} & = \frac{\displaystyle \partial \frac{\mu}{4 \pi} \int\limits_{\Omega '}  \int\limits_\Omega  \mathbf{J} \left( \mathbf{r} \right) \cdot \mathbf{J}^\ast \left( \mathbf{r '} \right) \frac{\mathrm{e}^{- \jmath k R}}{R} \, \mathrm{d} \mathbf{r} \, \mathrm{d} \mathbf{r '}}{\displaystyle \partial \omega} \\
& = \frac{k^2}{\omega} \left( \mathcal{L}_\omega \left( \mathbf{J},\mathbf{J} \right) - \jmath k \mathcal{L}_\mathrm{rad} \left( \mathbf{J},\mathbf{J} \right) \right)
\end{split}
\end{equation}
and
\begin{equation}
\label{app_b3}
\begin{split}
& \displaystyle\frac{\partial \mathcal{L} \left( \nabla \cdot \mathbf{J},\nabla \cdot \mathbf{J} \right)}{\partial \omega} \\
&= \frac{\displaystyle\partial \frac{1}{4 \pi\epsilon \omega^2} \int\limits_{\Omega '} \int\limits_\Omega \nabla \cdot \mathbf{J} \left( \mathbf{r} \right) \nabla \cdot \mathbf{J}^\ast \left( \mathbf{r '} \right) \frac{\mathrm{e}^{ -\jmath k R}}{R} \, \mathrm{d} \mathbf{r} \, \mathrm{d}\mathbf{r '}}{\partial \omega} \\
& = - \frac{1}{\omega} \Big( 2 \mathcal{L} \left( \nabla\cdot\mathbf{J}, \nabla\cdot\mathbf{J} \right) - \mathcal{L}_\omega \left( \nabla\cdot\mathbf{J}, \nabla\cdot\mathbf{J}\right) \\
& \qquad\qquad\qquad + \jmath k \mathcal{L}_\mathrm{rad} \left( \nabla\cdot\mathbf{J}, \nabla\cdot\mathbf{J} \right) \Big).
\end{split}
\end{equation}
It is important to remember that although the input current is normalized ($I_0 = 1$A), the current density $\mathbf{J}$ is still a function of angular frequency.

Putting all together and using the abbreviations (\ref{eqX6a}), (\ref{eqX6b}), (\ref{eqX9a}) and (\ref{eqX9b}) we immediately arrive at (\ref{eqX8}).

\section*{Acknowledgement}
The authors would like to thank Prof. Guy~A.~E.~Vandenbosch for fruitful discussions. Also, we would like to thank three anonymous reviewers whose remarks improved the clarity of the paper.

\ifCLASSOPTIONcaptionsoff
  \newpage
\fi

\bibliographystyle{IEEEtran}
\bibliography{references_LIST}

\begin{thebibliography}{10}
\providecommand{\url}[1]{#1}
\csname url@samestyle\endcsname
\providecommand{\newblock}{\relax}
\providecommand{\bibinfo}[2]{#2}
\providecommand{\BIBentrySTDinterwordspacing}{\spaceskip=0pt\relax}
\providecommand{\BIBentryALTinterwordstretchfactor}{4}
\providecommand{\BIBentryALTinterwordspacing}{\spaceskip=\fontdimen2\font plus
\BIBentryALTinterwordstretchfactor\fontdimen3\font minus
  \fontdimen4\font\relax}
\providecommand{\BIBforeignlanguage}[2]{{%
\expandafter\ifx\csname l@#1\endcsname\relax
\typeout{** WARNING: IEEEtran.bst: No hyphenation pattern has been}%
\typeout{** loaded for the language `#1'. Using the pattern for}%
\typeout{** the default language instead.}%
\else
\language=\csname l@#1\endcsname
\fi
#2}}
\providecommand{\BIBdecl}{\relax}
\BIBdecl

\bibitem{VolakisChenFujimoto_SmallAntennas_MiniatrurizTechniques}
J.~L. Volakis, C.-C. Chen, and K.~Fujimoto, \emph{Survey of Small Antenna
  Theory}, 1st~ed., ser. Small Antennas: Miniaturization Techniques \&
  Applications.\hskip 1em plus 0.5em minus 0.4em\relax McGraw-Hill, 2010.

\bibitem{Vandenbosch_ReactiveEnergiesImpedanceAndQFactorOfRadiatingStructures}
G.~A.~E. Vandenbosch, ``Reactive energies, impedance, and {Q} factor of
  radiating structures,'' \emph{IEEE Trans. Antennas Propag.}, vol.~58, no.~4,
  pp. 1112--1127, Apr. 2010.

\bibitem{Gustaffson_StoredElectromagneticEnergy_arXiv}
\BIBentryALTinterwordspacing
M.~Gustafsson and B.~L.~G. Jonsson. Stored electromagnetic energy and antenna
  q. eprint arXiv: 1211.5521. [Online]. Available:
  \url{http://adsabs.harvard.edu/abs/2012arXiv1211.5521G}
\BIBentrySTDinterwordspacing

\bibitem{GustafssonCismasuJonsson_PhysicalBoundsAndOptimalCurrentsOnAntennas_TAP}
M.~Gustafsson, M.~Cismasu, and B.~L.~G. Jonsson, ``Physical bounds and optimal
  currents on antennas,'' \emph{IEEE Trans. Antennas Propag.}, vol.~60, no.~6,
  pp. 2672--2681, June 2012, 

\bibitem{Rhodes_ObservableStoredEnergiesOfElectromagneticSystems}
D.~R. Rhodes, ``Observable stored energies of electromagnetic systems,''
  \emph{J. Franklin Inst.}, vol. 302, no.~3, pp. 225--237, 1976.

\bibitem{Rhodes_AReactanceTheorem}
------, ``A reactance theorem,'' \emph{Proc. R. Soc. Lond. A.}, vol. 353, pp.
  1--10, Feb. 1977, 

\bibitem{Rhodes_OnTheStoredEnergyOfPlanarApertures}
------, ``On the stored energy of planar apertures,'' \emph{IEEE Trans.
  Antennas Propag.}, vol.~14, no.~6, pp. 676--684, Nov. 1966, 

\bibitem{YaghjianBest_ImpedanceBandwidthAndQOfAntennas}
A.~D. Yaghjian and S.~R. Best, ``Impedance, bandwidth and {Q} of antennas,''
  \emph{IEEE Trans. Antennas Propag.}, vol.~53, no.~4, pp. 1298--1324, April
  2005, 

\bibitem{Jackson_ClassicalElectrodynamics}
J.~D. Jackson, \emph{Classical Electrodynamics}, 3rd~ed.\hskip 1em plus 0.5em
  minus 0.4em\relax John Wiley, 1998.

\bibitem{Carpenter_ElectromagneticEnergyAndPowerInTermsOfChargesAndPotentialsInsteadOfFields}
C.~J. Carpenter, ``Electromagnetic energy and power in terms of charges and
  potentials instead of fields,'' \emph{Proc. of IEE A}, vol. 136, no.~2, pp.
  55--65, March 1989.

\bibitem{HazdraCapekEichler_CommentsToGuy1}
P.~Hazdra, M.~Capek, and J.~Eichler, ``Comments to "reactive energies,
  impedance, and {Q} factor of radiating structures" by {G. Vandenbosch},''
  \emph{IEEE Trans. Antennas Propag.}, vol.~xx, p.~xx, 2013, after Major rev.

\bibitem{HarringtonMautz_TheoryOfCharacteristicModesForConductingBodies}
R.~F. Harrington and J.~R. Mautz, ``Theory of characteristic modes for
  conducting bodies,'' \emph{IEEE Trans. Antennas Propag.}, vol.~19, no.~5, pp.
  622--628, Sept. 1971.

\bibitem{CapekHazdraEichler_AMethodForTheEvaluationOfRadiationQBasedOnModalApproach}
M.~Capek, P.~Hazdra, and J.~Eichler, ``A method for the evaluation of radiation
  {Q} based on modal approach,'' \emph{IEEE Trans. Antennas Propag.}, vol.~60,
  no.~10, pp. 4556--4567, Oct. 2012.

\bibitem{feko}
\BIBentryALTinterwordspacing
EM Software \& Systems-S.A. FEKO. [Online]. Available: \url{www.feko.info}
\BIBentrySTDinterwordspacing

\bibitem{cst}
\BIBentryALTinterwordspacing
CST Computer Simulation Technology. [Online]. Available: \url{www.cst.com}
\BIBentrySTDinterwordspacing

\bibitem{GustafssonNordebo_BandwidthQFactorAndResonanceModelsOfAntennas}
M.~Gustafsson and S.~Nordebo, ``Bandwidth, {Q} factor and resonance models of
  antennas,'' \emph{PIER}, vol.~62, pp. 1--20, 2006, 

\bibitem{Harrington_TimeHarmonicElmagField}
R.~F. Harrington, \emph{Time-Harmonic Electromagnetic Fields}, 2nd~ed.\hskip
  1em plus 0.5em minus 0.4em\relax John Wiley - IEEE Press, 2001.

\bibitem{CollinRotchild_EvaluationOfAntennaQ}
R.~E. Collin and S.~Rotchild, ``Evaluation of antenna {Q},'' \emph{IEEE Trans.
  Antennas Propag.}, vol.~12, pp. 23--27, 1964, 

\bibitem{Harrington_FieldComputationByMoM}
R.~F. Harrington, \emph{Field Computation by Moment Methods}.\hskip 1em plus
  0.5em minus 0.4em\relax John Wiley - IEEE Press, 1993.

\bibitem{ArcioniBressanPerregrini_OnTheEvaluationOfTheDoubleSurfaceIntegralsArisingInTheApplicationOfETC}
P.~Arcioni, M.~Bressan, and L.~Perregrini, ``On the evaluation of the double
  surface integrals arising in the application of the boundary integral method
  to 3-{D} problems,'' \emph{IEEE Trans. Microwave Theory Tech.}, vol.~44,
  no.~3, pp. 436--438, March 1997.

\bibitem{CapekHamouzHazdraEichler_ImplementationOfTCMinMatlab}
M.~Capek, P.~Hamouz, P.~Hazdra, and J.~Eichler, ``Implementation of the {Theory
  of Characteristic Modes in Matlab},'' \emph{IEEE Antennas Propag. Magazine},
  vol.~55, no.~2, pp. 176--189, April 2013.

\bibitem{RaoWiltonGlisson_ElectromagneticScatteringBySurfacesOfArbitraryShape}
S.~M. Rao, D.~R. Wilton, and A.~W. Glisson, ``Electromagnetic scattering by
  surfaces of arbitrary shape,'' \emph{IEEE Trans. Antennas Propag.}, vol.~30,
  no.~3, pp. 409--418, May 1982.

\bibitem{matlab}
\BIBentryALTinterwordspacing
{The MathWorks}. The Matlab. [Online]. Available: \url{www.mathworks.com}
\BIBentrySTDinterwordspacing

\bibitem{CapekHazdraEichlerHamouzMazanek_AccelerationTechniquesInMatlab}
M.~Capek, P.~Hazdra, J.~Eichler, P.~Hamouz, and M.~Mazanek, ``Acceleration
  techniques in {M}atlab for {EM} community,'' in \emph{Proceedings of the 7th
  European Conference on Antennas and Propagation (EUCAP)}, Gothenburg, Sweden,
  April 2013.

\bibitem{Balanis_AdvancedTheory}
C.~A. Balanis, \emph{Advanced Engineering Electromagnetics}.\hskip 1em plus
  0.5em minus 0.4em\relax John Wiley, 1989.

\end{thebibliography}

\begin{biography}[{\includegraphics[width=1in,height=1.25in,clip,keepaspectratio]{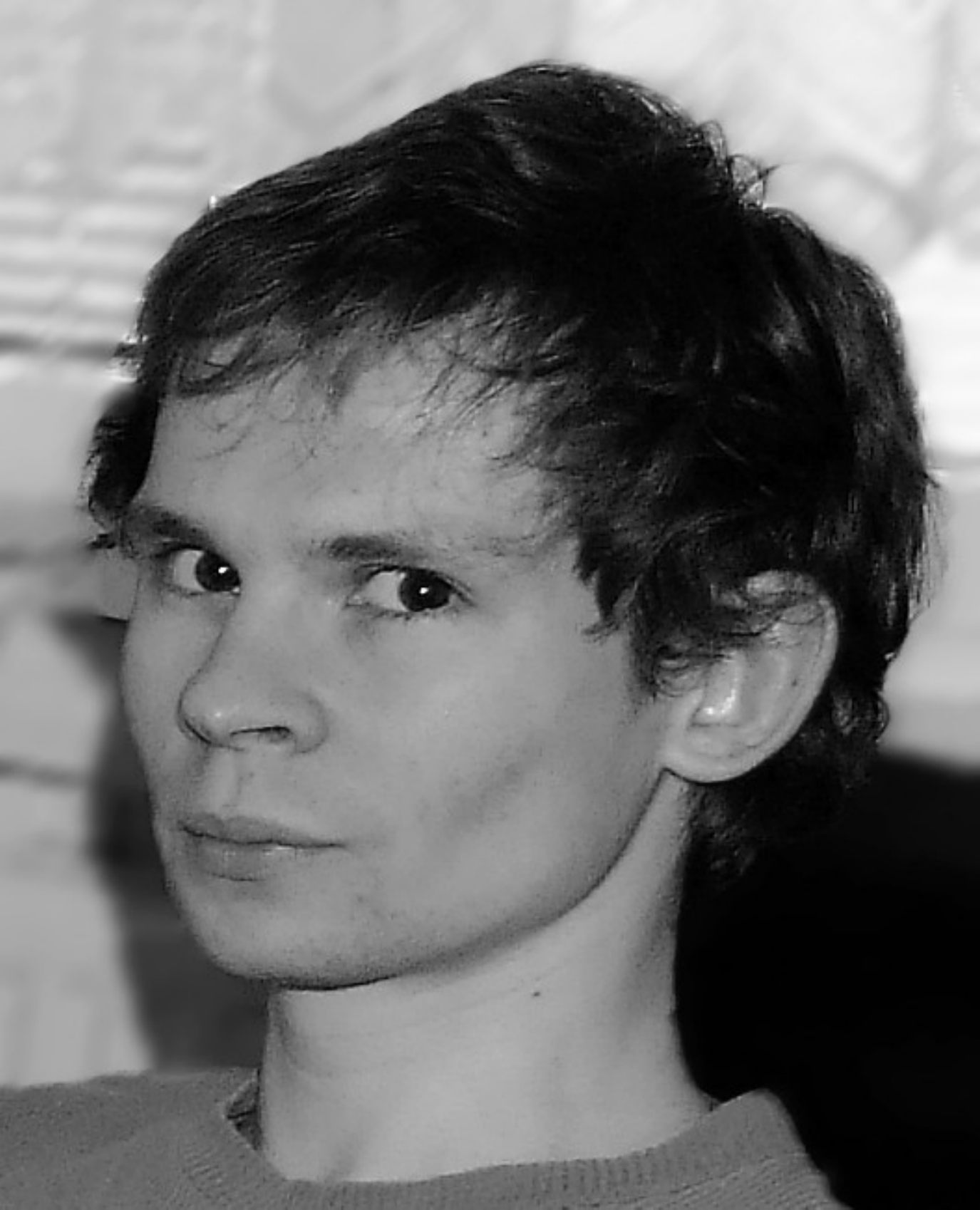}}]{Miloslav Capek}
(S'09) received the M.Sc. degree in electrical engineering from the Czech Technical University, Prague, Czech Republic, in 2009, and is currently working towards a Ph.D. degree in
electromagnetic fields at the same University.

His research interests are in the area of electromagnetic theory, electrically small antennas, numerical techniques, fractal geometry and optimization.

\end{biography}
\begin{biography}[{\includegraphics[width=1in,height=1.25in,clip,keepaspectratio]{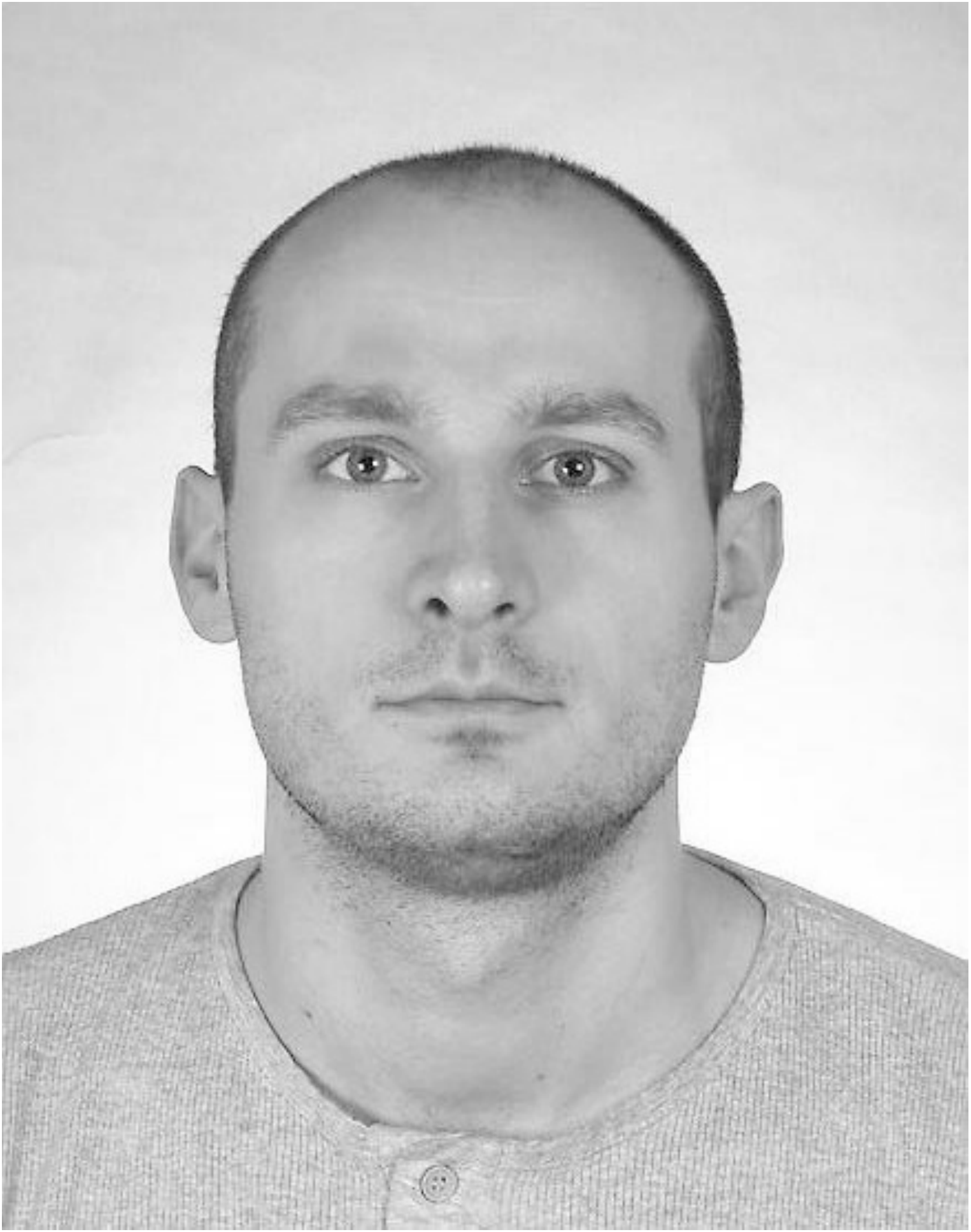}}]{Lukas Jelinek}
received the Ph.D. degree from the Czech Technical University in Prague, Prague, Czech Republic, in 2006.

Currently, he is a researcher with the Department of Electromagnetic Field, CTU-FEE.  His main fields of interest include wave propagation in complex media, general field theory, and numerical techniques. His recent research interest is focused on metamaterials, specifically on resonant ring systems.
\end{biography}

\begin{biography}[{\includegraphics[width=1in,height=1.25in,clip,keepaspectratio]{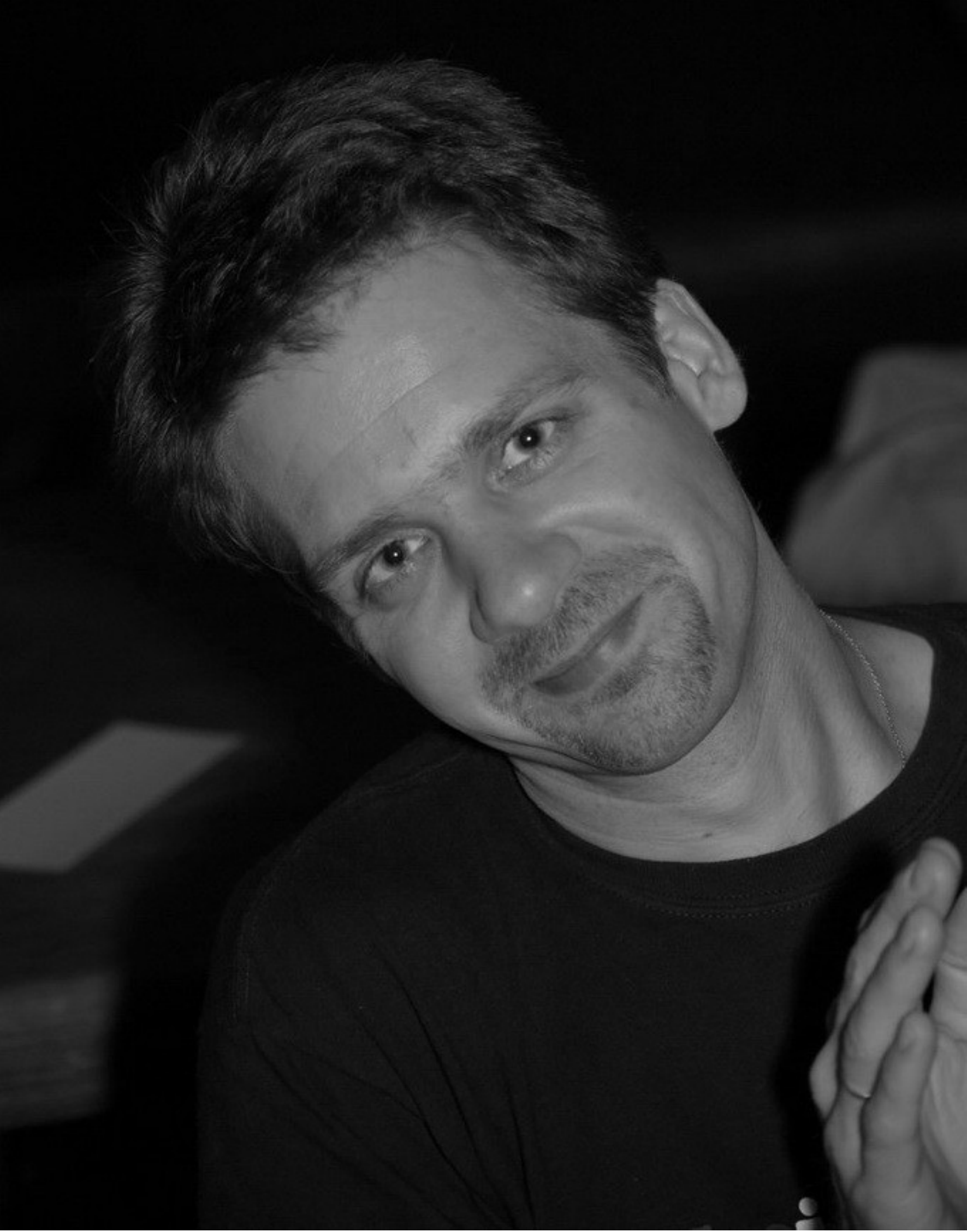}}]{Pavel Hazdra}
(M'03) received the M.S. and Ph.D. degrees in electrical engineering from the Czech Technical University in Prague, Faculty of Electrical Engineering in 2003 and 2009, respectively.

He is a research and teaching assistant with the Department of Electromagnetic Field, CTU-FEE. His research interests are in the area of electromagnetic theory, computational electromagnetics, fractal geometry, planar antennas and special prime-feed antennas.

\end{biography}
\begin{biography}[{\includegraphics[width=1in,height=1.25in,clip,keepaspectratio]{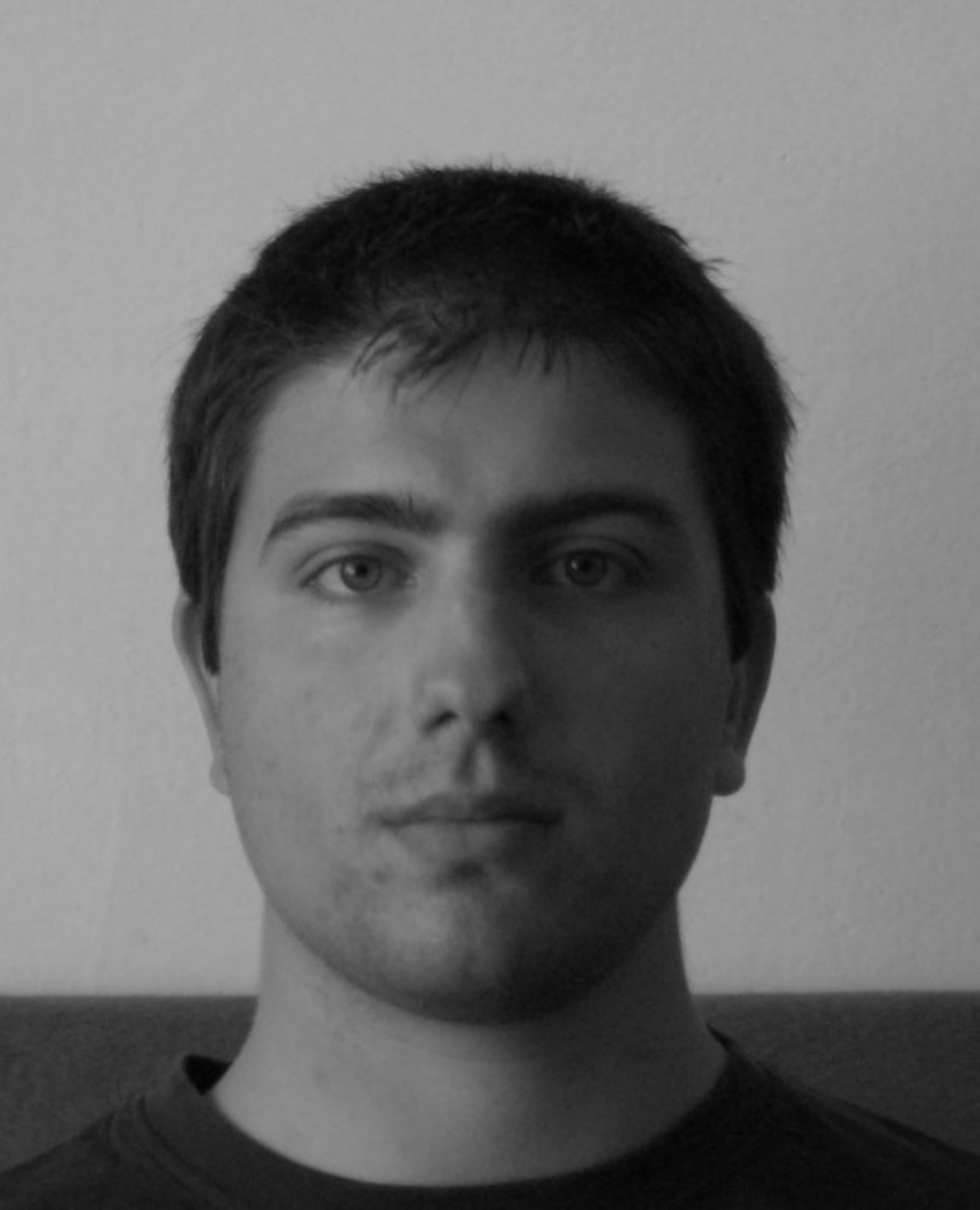}}]{Jan Eichler}
(S'10) received the B.Sc. and M.Sc. degrees in electrical engineering from the Czech Technical University in Prague in 2008 and 2010, respectively. He is currently working towards a Ph.D.
degree at the same University.

His research interests include modal methods for antenna design and connecting them with full-wave methods. He is also interested in developing and simulating active antennas.
\end{biography}
\end{document}